\documentclass[runningheads]{llncs}
\usepackage{latexsym,amsfonts,amssymb}
\input{epsf}
\newenvironment{Figure}[1][t]{%
\begin{figure}[#1]\begin{minipage}{\textwidth}}%
{\end{minipage}\end{figure}}
\newcommand{\Chop}{{\downarrow}}
\newcommand{\And}{\land}
\newcommand{\Registered}{{\bigcirc\hspace{-.5em}r}}
\newcommand{\Ie}{\textit{i.e.}}
\newcommand{\Const}[1]{\underline{#1}}
\newcommand{\QTL}{\textit{qM$\mu$}}
\newcommand{\QMU}{\textit{qM$\mu$}}
\newcommand{\Mins}{{\underline{\sigma}}}
\newcommand{\Maxs}{{\overline{\sigma}}}
\newcommand{\Tree}[3]{[\![ #1 ]\!] _{#2} ^{#3}}
\newcommand{\Val}{\textit{Val}}
\newcommand{\SmallExp}{\raisebox{0em}[0em][0em]{\small$\int$}}
\newcommand{\Exp}[1]{\int_\protect
    {\makebox[0pt][l]{\raisebox{-1.2ex}{\scriptsize $#1$}}}~}
\newcommand{\EXP}[1]{\int_\protect
    {\makebox[0pt][l]{\raisebox{-2ex}{\scriptsize $#1$}}}~}
\newcommand{\Koz}[2]{|\!| #1 |\!| _{#2}}
\newcommand{\KozS}[3]{|\!|\!| #1 |\!|\!| _{#2} ^{#3}}
\newcommand{\VV}{{\cal V}}
\newcommand{\LHS}{\textit{lhs}}
\newcommand{\RHS}{\textit{rhs}}

\newcommand{\Times}{\times}
\newcommand{\SDollar}{S_\$}

\newcommand{\NextTime}[1]{\{#1\}}
\newcommand{\Concat}{\mathop{{+}\!\!\!{+}}}
\newcommand{\PR}[1]{{\cal R}.#1}
\newcommand{\E}[1]{{\cal E}#1}
\newcommand{\Dem}[1]{[#1]}
\newcommand{\Ang}[1]{\langle #1 \rangle}
\newcommand{\Bchoice}[3]{#2 \lhd #1 \rhd #3}
\newcommand{\PPle}{\leq}
\newcommand{\KKK}{\textsf{K}}
\newcommand{\GGG}{\textsf{G}}
\newcommand{\AAA}{\textsf{A}}
\newcommand{\CCC}{{\sf C}} 
\newcommand{\KK}{\textsf{k}}
\newcommand{\MIN}{\textit{Min}}
\newcommand{\MAX}{\textit{Max}}
\newcommand{\UnderB}{\underline{\sf G}}
\newcommand{\OverB}{\overline{\sf G}}

\newcommand{\PTL}{\textit{qTL}}
\newcommand{\EG}{\textit{e.g.}}
\newcommand{\IE}{\textit{i.e.}}

\newcommand{\EtAl}{\textit{et.\ al.}}
\newcommand{\OK}{\textsf{ok}}
\newcommand{\True}{\textsf{true}}
\newcommand{\False}{\textsf{false}}
\newcommand{\PC}[1]{\mathbin{\makebox[0em]{~}_{#1}\oplus}}
\newcommand{\SetCompYN}[2]{\{#1 \mid #2\}}
\newcommand{\CompNY}[3]{(#1#2 \Spot #3)}
\newcommand{\Spot}{\mathrel{\mbox{\boldmath$\cdot$}}}
\newcommand{\Lift}[1]{\overline{#1}}
\newcommand{\In}{\colon}
\newcommand{\Defs}{\mathrel{\hat{=}}}
\newcommand{\Lle}{\ll}
\newcommand{\Min}{\sqcap}
\newcommand{\Cond}[3]{#1~\textrm{\underline{if}}~#2~\textrm{\underline{else}}~#3}

\newcommand{\Sup}{\sqcup}
\newcommand{\Inf}{\sqcap}
\newcommand{\Lfp}{\mathop{\sf lfp}}
\newcommand{\Gfp}{\mathop{\sf gfp}}

\newcommand{\Max}{\sqcup}
\newcommand{\Implies}{\mathrel{\Rightarrow}}
\newcommand{\Wide}[1]{~~~#1~~~}

\newcommand{\Sec}[1]{Sec.~\ref{#1}}
\newcommand{\App}[1]{App.~\ref{#1}}
\newcommand{\Fig}[1]{Fig.~\ref{#1}}
\newcommand{\Eqn}[1]{(\ref{#1})}
\newcommand{\Lem}[1]{Lem.~\ref{#1}}
\newcommand{\Def}[1]{Def.~\ref{#1}}
\newcommand{\Thm}[1]{Thm.~\ref{#1}}
\newenvironment{Reason}{\begin{tabbing}\hspace{4em}\=\kill}{\end{tabbing}\vspace{-2.5ex}}
\newcommand{\Space}{~ \\}
\newcommand{\Step}[2]{#1 \> $\begin{array}[t]{ll}#2\end{array}$ \\}
\newcommand{\StepR}[3]{#1 \> $\begin{array}[t]{ll}#3\end{array}$ \` {\RF #2} \\}

\newcommand{\WideStepR}[3]{#1 \` {\RF #2} \\ \> $\begin{array}[t]{ll}#3\end{array}$ \\}
\newcommand{\RF}{\small}
\newcommand{\Gets}{\colon=\,}

\begin{document}
\mainmatter
\title{Results on the quantitative $\mu$-calculus \QMU%
\thanks{An earlier and much-abridged version appeared in \emph{Proc.~LPAR 2002, Tbilisi} \cite{GPatQMCQ}.}}
\titlerunning{\texttt{15 September 2003} \hspace{5em} Results on the quantitative $\mu$-calculus \QMU}
\author{AK McIver\inst{1} \and CC Morgan\inst{2}}
\authorrunning{AK McIver and CC Morgan}
\tocauthor{AK McIver (Macquarie University) and CC Morgan (University of New South Wales)}
\institute{Dept.~Computer Science,
Macquarie University,
NSW 2109 Australia;
\email{anabel@ics.mq.edu.au}
\and
Dept.~Comp.~Sci.~\&~Eng.,
University of New South Wales,
NSW 2052 Australia;
\email{carrollm@cse.unsw.edu.au}}
%
\maketitle
\begin{abstract}
The $\mu$-calculus is a powerful tool for specifying and
verifying transition systems, including those with both demonic (universal) and
angelic (existential) choice; its \emph{quantitative} generalisation \QMU\ \cite{QAaMC,aPTCboE,QSoORG} extends that
to \emph{probabilistic} choice.

We show here that for a finite-state system the \emph{logical} interpretation of \QMU, \emph{via} fixed-points in a domain of real-valued functions into $[0,1]$, is equivalent to an \emph{operational}
interpretation given as a turn-based gambling game between two
players.

\bigskip
The equivalence sets \QMU\ on a par with the standard $\mu$-calculus,
in that it too can benefit from a solid interface linking the
logical and operational frameworks.

The logical interpretation provides direct access to axioms, laws and meta-theorems.
The operational, game- based interpretation aids the intuition and continues in the
more general context to provide a surprisingly practical
specification tool --- meeting for example Vardi's challenge to
``figure out the meaning of
$\textsf{AF}\,\textsf{AX}\,p$" as a branching-time formula.

\bigskip
A corollary of our proofs is an extension of Everett's singly-nested games result in the finite turn-based case: we prove well-definedness of the \emph{minimax} value, and existence of fixed memoriless strategies, for all \QMU\ games/formulae, of arbitrary (including alternating) nesting structure.
\end{abstract}

\section{Introduction}
\label{sec1}

The standard $\mu$-calculus, introduced by Kozen \cite{RotPMC},
extends Boolean dynamic program logic by the introduction of least
($\mu$) and greatest ($\nu$) fixed-point operators. Its proof
system is applicable to both infinite and finite state spaces;
recent results \cite{NotPMCCaRR} have established a complete
axiomatisation; and it can be specialised to temporal logic.
Thus it has a simple semantics, and a proof theory.

But its operational significance can be more elusive:
general $\mu$-calculus expressions can be difficult to use, because
in all but the simplest cases they are not easy on the intuition. Alternating fixed points can be especially intricate; and even the more straightforward (alternation-free) temporal subset has
properties (particularly ``branching-time properties") that are
notoriously difficult to specify, as Vardi points out \cite{BvLTFS}. 

\bigskip
Stirling's
``two-player-game'' interpretation alleviates this problem by providing an alternative and complementary
operational view \cite{LMCG}.

\bigskip
The \emph{quantitative} modal $\mu$-calculus acts over
\emph{probabilistic} transition systems, extending the standard axioms
from Boolean- to real values; and it
would benefit just as much from having two complementary
interpretations. Our goal in this paper is to identify them, and to give the proof of their equivalence, over a finite state space: one
interpretation (defined earlier \cite{aPTCboE,QAaMC,QSoORG}) generalises Kozen's by lifting it from the Booleans into the reals; the other (defined here)
generalises Stirling's.

Our principal contribution here is thus the definition of the Stirling-style but \emph{quantitative} interpretation, and the proof of its equivalence to the Kozen-style quantitative interpretation.%
\footnote{The quantitative Kozen interpretation has been given earlier \cite{QAaMC,aPTCboE,QSoORG}; we review it here.}
We show also that memoriless
strategies suffice, for
both interpretations, again when the state space is finite.

\bigskip The Kozen-style quantitative interpretation is based on our earlier extension \cite{aPPDL,PPT} of
Dijkstra/Hoare logic to probabilistic/demonic programs (corresponding to the $\forall$ modality): it is a real-valued logic based on ``greatest
pre-expectations'' of random variables, rather than weakest
preconditions of predicates. It can express the specific ``probability of
achieving a postcondition,'' since the probability of an event is
the expected value of its characteristic function,%
\footnote{\label{n1135}See \Sec{s1135} for an example of this.}
but it applies
more generally to other cost-based properties besides. Although the
specific approach, \IE\ with its explicit probabilities --- may be  more intuitive, the extra generality of a full quantitative logic seems necessary for compositionality \cite{PHHP}.

Converting predicates ``wholesale'' from Boolean- to real-valued
state functions --- due originally to Kozen \cite{SoPP} and extended by us to
include demonic (universal) \cite{PPT} and angelic (existential) \cite{DAaUPCiSP}
nondeterminism --- contrasts with probabilistic logics using
``threshold functions'' \cite{MCoPaNS,PTLvtMMC} that mix Boolean
and numeric arguments: the uniformity in our case means that standard Boolean 
identities in branching-time temporal logic \cite{tTLoBT} suggest corresponding quantitative laws for us \cite{aEBMfPTL}, and so we get a
powerful collection of algebraic properties ``for free.'' The
logical ``implies" relation between Booleans is replaced by the
standard ``$\leq$'' order on the reals; $\False$ and $\True$ become 0 and 1;
and fixed points are then
associated with monotonic \emph{real}- rather than Boolean-valued functions. The resulting arithmetic logic is applicable to a restricted class of
real-valued functions, and we recall its definition in \Sec{sec3}.

\bigskip
Our Stirling-style quantitative interpretation is operational, and is based on
his earlier strategy-based game metaphor for the standard
$\mu$-calculus. In our richer context, however, we must distinguish
nondeterministic choice --- both demonic and angelic --- from probabilistic choice: the former continues to be represented by the two
players' strategies; but the latter is represented by the new
feature that we
make the players \emph{gamble}. In \Sec{game} we set out the details.

\bigskip
In \Sec{sec5} we give a worked
example
\footnote{using both \emph{PRISM} \cite{KNP02a} and \emph{Mathematica}$^\Registered$}
of the full use of the quantitative aspects of the calculus, beyond simply calculating probabilities.%

\bigskip
The main mathematical result of this paper is given in \Sec{dooalidy}, though much of the detail is placed in the appendices.

Stirling showed that for non-probabilistic formulae the
Boolean value of the Kozen interpretation corresponds to the
existence of a winning strategy in his game interpretation. In
our case, strategies in the game must become
``optimal'' rather than ``winning''; and the correspondence is
now between a formula's value (since it denotes a real number, in the
Kozen interpretation) and the expected winnings from the
zero-sum gambling game (of the Stirling interpretation).
Since the
gambling game described by a formula is a ``\emph{minimax},'' we must show it to be well-defined (equal to the ``\emph{maximin}''): in fact we show that both the \emph{minimax} and the \emph{maximin} of the game are equal to the Kozen-style denotation of the formula that generated it.

We also prove that memoriless strategies
suffice.

Both proofs apply only to finite state spaces.

\bigskip
The benefit of our proved equivalence is to set the quantitative
$\mu$-calculus on a par with standard $\mu$-calculus in that a
suitable form of ``logical validity" corresponds exactly to an
operational interpretation. As with standard $\mu$-calculus, a
specifier can use the \emph{operational} semantics to build his intuitions into a game, and can then use the
general features of the logic --- whose soundness has been proved relative to the \emph{logical}%
\footnote{We also call this the \emph{denotational} semantics.}
semantics --- to prove properties about the specific application. For
example, the \emph{sublinearity} \cite{PPT} of \QTL\ --- the quantitative
generalisation of the \emph{conjunctivity} of standard modal algebras --- has
been used
in its quantitative temporal subset \PTL\ to prove a number of
algebraic laws corresponding to those holding in standard
branching-time temporal logic \cite{aEBMfPTL}.

Preliminary experiments have shown that the proof system is very
effective for unravelling the intricacies of distributed protocols
\cite{tCCP,pGCL}. Moreover it provides an attractive proof
framework for Markov decision processes \cite{CAoGuPL,CMDP} --- and
indeed many of the problems there have a succinct specification as
$\mu$-calculus formulae, as the example of \Sec{LSM} illustrates.
In ``reachability-style problems" \cite{CMaMRTiPS}, proof-theoretic
methods based on the logic presented here have produced very direct
arguments related to the abstraction of probabilities
\cite{ACEaAPiQTL}, and even more telling is that the logic is
applicable even in infinite state spaces \cite{CMaMRTiPS}. All of
which is to suggest that further exploration of $\QTL$ will
continue to be fruitful.

\bigskip
In the following we shall assume generally that $S$ is a countable
state space (though for the principal result we restrict to
finiteness, in \Sec{dooalidy}). If $f$ is a function with domain
$X$ then by $f.x$ we mean $f$ applied to $x$, and $f.x.y$ is
$(f.x).y$ where appropriate; functional composition is written with \mbox{\scriptsize$\circ$}, so that $(f\mbox{\scriptsize$\circ$}g).x = f.(g.x)$.
We denote the set of discrete
probability \emph{sub}-distributions over a set $X$ by $\Lift{X}$: it is the set of functions
from $X$ into the real interval $[0,1]$ that sum to \emph{no more than} one; and
if $A$ is a random variable with respect to some probability space, and
$\delta$ is some probability sub-distribution, we write $\int_\delta A$
for the expected value of $A$ with respect to $\delta$.%
\footnote{Normal mathematical practice is to write $\int A\, \textrm{d}\delta$,
but that greatly confuses the roles of bound and free variables: it makes the distribution (measure) variable $\delta$ in $\textrm{d}\delta$ free in the expression; but in the analogous $\int f(x) \textrm{d}x$ of analysis, the independent variable $x$ in $\textrm{d}x$ is bound.}
In the special case that $\delta$ is in $\Lift{X}$ and $A$ is a bounded
real-valued function on $X$, in fact $\int_\delta A$ is equal to $\sum_{s:S}
A.s \times \delta.s$.

\section{Probabilistic transition systems and $\mu$-calculus}
\label{language}\label{sec2}

In this section we set out the logical language, together with some details about the
probabilistic systems over which the formulae are to be interpreted.

Formulae in the logic (in positive%
\footnote{The restriction to the positive fragment is for the usual reason: that
the interpretation of any expression $\CompNY{\lambda}{X}{\phi}$, constructed
according to the given rules, should yield a monotone function of $X$.}
form) are constructed as follows:
\[
 \phi
 \Wide{\Defs}
    X \mid
    \AAA \mid
    \Ang{\KKK} \phi \mid
    \Dem{\KKK}\phi \mid
    \phi_1 \Min \phi_2 \mid
    \phi_1 \Max \phi_2 \mid
    \Bchoice{\GGG}{\phi_1}{\phi_2} \mid
    \CompNY{\mu}{X}{\phi} \mid
    \CompNY{\nu}{X}{\phi} ~.
\]
\begin{itemize}
\item Variables $X$ are of type $S \rightarrow [0,1]$, and are used for binding fixed points.
\item Terms $\AAA$ stand for fixed functions in $S \rightarrow [0,1]$.
\item Terms $\KKK$ represent finite non-empty sets of probabilistic state-to-state transitions in $\PR{S}$ (see below), with $\Ang{\cdot}$ and $\Dem{\cdot}$ forming respectively angelic- (existential-) and demonic (universal) modalities from them.
\item Terms $\GGG$ describe Boolean functions of $S$, used in 
\(
 \lhd~\textrm{(``if'')}~\GGG~\rhd~\textrm{(``else'')}
\)
style \cite{aCoNitPC}.
\end{itemize}

It is well known that such formulae can be used to express complex
path-properties of computational sequences. In this paper we  interpret
the formulae over sequences based on generalised probabilistic transitions%
\footnote{They correspond to the ``game rounds'' of Everett \cite{RG}.}
in what we call $\PR{S}$, the functions $t$
in $S \rightarrow \Lift{\SDollar}$ where $\SDollar$ is just the state space $S$ with a special ``payoff''
state $\$$ adjoined. Thus $\Lift{\SDollar}$ is the set of sub-distributions over that, so that the elements $t$ of $\PR{S}$ give
the probability of passage from initial $s$ to final (proper) $s'$ 
as $t.s.s'$; any deficit $1-\sum_{s'}t.s.s'$ is interpreted as the probability
of an immediate halt with payoff
\begin{equation}\label{e0946}
t.s.\$/(1-\sum_{s'\In S}t.s.s') ~. \footnotemark
\end{equation}
See \Fig{f1154} for an example.

This
\footnotetext{If $\sum_{s'\In S}t.s.s'$ is one, then $t.s.\$$ must be zero, because elements of $\Lift{\SDollar}$ sum to no more than one. In that case we define the actual --- and expected --- payoffs both to be zero.}
formulation of the payoff --- \IE\ ``pre-divided'' by its probability of occurrence --- has three desirable properties. The first is
that the probabilistically \emph{expected} halt-and-payoff is just $t.s.\$$, \IE\ is given directly by $t$.
The second property is
that we can consider the probabilities of outcomes from $s$ to sum to one exactly
(rather than no more than one), since any deficit is ``soaked up'' in the
probability of transit to payoff, which simplifies our operational
interpretation.
\begin{Figure}

The relational element
\raisebox{0pt}[3\baselineskip][0pt]{
\(\left(\begin{array}{rcl}
 s &\stackrel{1/4}{\longrightarrow}& \textsf{H} \\
 s &\stackrel{1/4}{\longrightarrow}& \textsf{T} \\
 s &\stackrel{2/5}{\longrightarrow}& \textrm{\$}
\end{array}\right)
\)
}
in $\PR{S}$, with its deficit of $1/10$, denotes the transition shown,
\begin{quote}
\setlength{\unitlength}{1mm}
\begin{picture}(40,22)(-60,18)
\put(14,32){\makebox(0,0){$s$}}
\put(16,32){\line(1,1){7}}
\put(30,41){\vector(1,0){3}}
\put(24,41){\makebox(0,0)[l]{\scriptsize $1/4$}}
\put(34,41){\makebox(0,0)[l]{\textsf{H$\,\cdots$}}}
\put(16,32){\line(1,0){7}}
\put(30,32){\vector(1,0){3}}
\put(24,32){\makebox(0,0)[l]{\scriptsize $1/4$}}
\put(34,32){\makebox(0,0)[l]{\textsf{T$\,\cdots$}}}
\put(16,32){\line(1,-1){7}}
\put(30,22){\vector(1,0){3}}
\put(24,22){\makebox(0,0)[l]{\scriptsize $1/2$}}
\put(34,22){\makebox(0,0)[l]{$\$0.80$}}
\end{picture}
\end{quote}
\vspace{-7.5em}
\parbox[t]{0.55\textwidth}{
in which the transition probabilities now sum to one. In particular, the probability of transition to $\$$ is $1/2$, which makes the \emph{expected} immediate payoff equal to
$1/2 \times 0.80 = 2/5$ as given explicitly in the relation.}

\bigskip
Since $\textsf{H},\textsf{T}$ are states, they may lead further:
no matter where they lead, however, the expected reward of the subtrees rooted there cannot exceed one, and so our encoding ensures both that the transition probabilities from $s$ sum to one exactly (since the probability of transition $s \mapsto \$$ is $1/2 = 1 - (1/4 + 1/4)$ by definition), and that the expected reward from this tree (rooted at $s$) cannot exceed one either (since the actual reward $\$0.80$ is defined just so that $2/5 = 0.8 \times 1/2$ will hold).

\bigskip
The tree does not continue on from the payoff state $\$0.80$.

\bigskip
More generally, a ``normal'' transition, \Ie\ with $\sum_i p_i = 1$, can effectively be ``$\alpha$-discounted'' by using the elements $(s\stackrel{\alpha p_i~}{\longrightarrow} s_i, s\stackrel{0}{\longrightarrow} \$)$ or $(s\stackrel{\alpha p_i~}{\longrightarrow} s_i, s\stackrel{1{-}\alpha}{\longrightarrow} \$)$.

\caption{\label{f1154} Example of payoff-state encoding}
\end{Figure}

The third property is that transitions preserve one-boundedness in the following sense.
Define the set of expectations $\E{S}$ (over $S$) to
be the set of one-bounded functions $S \rightarrow [0,1]$. If $A$ in
$\E{S}$ gives a ``post-expectation'' $A.s'$ expected to be realised at state $s'$
after transition $t$, then the ``pre-expectation'' at $s$ before transition $t$ is
\[
	t.s.\$ + \int_{t.s} A ~,
	\quad \parbox[t]{20em}{where the sub-distribution $t.s$
	             under $\int$ is restricted to states in $S$ proper.%
	             \footnotemark}
\]
\footnotetext{To avoid clutter we will assume this restriction where necessary in the sequel.}%
It is the expected value realised by making transition $t$ from $s$ to $s'$ or
possibly $\$$, taking $A.s'$ in the former case and \Eqn{e0946} in the latter. That this pre-expectation
is also one-bounded, \IE\ is in $\E{S}$, allows us to confine our work to the real interval $[0,1]$
throughout.

Hence computation trees can be 
constructed by ``pasting together'' applications of transitions $t_0,
t_1,\ldots$
drawn from
$\PR{S}$, with branches to $\$$ being tips.%
\footnote{We see below that tips are made by constant terms $\AAA$ as well.}
The probabilities attached to the
individual steps then generate a
distribution over computational paths (which is defined by the sigma-algebra of
extensions of finite sequences, a well-known construction \cite{PaI}).

We use the relation $\PPle$ --- ``everywhere no more than" between
expectations (thus replacing ``implies"):
\[
 A \PPle A'
 \Wide{\textrm{iff}}
 \CompNY{\forall}{s\In S}{A.s \leq A'.s}~.
\]

In our interpretations we will use \emph{valuations} in the usual way. Given a
formula $\phi$, a valuation $\VV$ does four things: (i) it maps each $\AAA$
in $\phi$ to a fixed expectation in $\E{S}$; (ii) it maps each
$\KKK$ to a fixed, non-empty finite set of probabilistic transitions in $\PR{S}$; (iii) it
maps each $\GGG$ to a predicate over $S$; and (iv) it keeps track
of the current instances of ``unfoldings'' of fixed points, by including
mappings
for bound variables $X$. (For notational economy, in (iv) we are allowing $\VV$
to take over the role usually given to a separate ``environment'' parameter.)

We make one simplification to our language, without compromising expressivity. Because the valuation $\VV$
assigns \emph{finite} sets to all occurrences of $\KKK$, we can replace each
modality $\Ang{\KKK} \phi$ (resp.~$\Dem{\KKK}\phi$) by an explicit maxjunct
$\Max_{\KK \In \KKK}\NextTime{\KK}\phi$ (resp.~minjunct $\Min_{\KK \In \KKK}
\NextTime{\KK}\phi$) of (symbols $\KK$ denoting) transitions $k$ in the set
(denoted by) $\KKK$. We do this because our interpretations conveniently do not distinguish
between $\Ang{\KKK}$ or $\Dem{\KKK}$ when $\KKK$ is a singleton set.

In the rest of this paper we shall therefore use the \emph{reduced language}
given by
\[
 \phi
 \Wide{\Defs}
    X \mid
    \AAA \mid
    \NextTime{\KK}\phi \mid
    \phi_1 \Min \phi_2 \mid
    \phi_1 \Max \phi_2 \mid
    \Bchoice{\GGG}{ \phi_1}{ \phi_2} \mid
    \CompNY{\mu}{X}{\phi} \mid
    \CompNY{\nu}{X}{\phi} ~.
\]
We replace (ii) above in respect of $\VV$ by: (ii') it maps each occurrence of
$\NextTime{\KK}$ to a probabilistic transition in $\PR{S}$.

\section[Denotational interpretation]{\begin{tabular}[t]{@{}l}
          Denotational interpretation: \\
          \QMU\ generalises Kozen's logic
         \end{tabular}
         }
\label{qTL}\label{sec3}
 
In this section we recall how the quantitative logic for nondeterministic/probabilistic
sequential programs \cite{aPPDL,PPT}
--- from which we inherit the use of expectations, and the semantic definition 
$\Koz{\NextTime{\KK} \phi}{}$ below --- 
leads to a quantitative generalisation of Kozen's logical
interpretation of $\mu$-calculus, suitable for probabilistic transition systems.

Let $\phi$ be a formula and $\VV$ a valuation. We write $\Koz{\phi}{\VV}$ for
its meaning, an expectation in $\E{S}$ determined by the rules given in \Fig{f1210}.
Part of the contribution of our previous work \cite{aPTCboE,aEBMfPTL} is summarised in the following lemma.

\begin{Figure}
\begin{enumerate}
\item\label{i1506} $\Koz {X}{\VV} \Wide{\Defs} \VV.X$ ~.
\item $\Koz {\AAA}{\VV} \Wide{\Defs} \VV.\AAA$ ~.
\item\label{i0959}
    $\Koz{\NextTime{\KK} \phi}{\VV}.s
    \Wide{\Defs} \VV.\KK.s.\$ ~+~ \Exp{\VV.\KK.s}\Koz{\phi}{\VV}$ ~.
\item[] 
\item\label{i1219}
    $\Koz{\phi' \Min \phi'' }{\VV}.s
    \Wide{\Defs}
    \Koz{\phi'}{\VV}.s~\textsf{min}~\Koz{\phi''}{\VV}.s$ ~; and \newline 
    $\Koz{\phi' \Max \phi''}{\VV}.s
    \Wide{\Defs}
    \Koz{\phi'}{\VV}.s~\textsf{max}~\Koz{\phi''}{\VV}.s$ ~.
\item[]
\item
    $\Koz{\Bchoice{\GGG}{\phi'}{\phi''}}{\VV}.s
    \Wide{\Defs}
    \Cond{\Koz{\phi'}{\VV}.s}{(\VV.\GGG.s)}{\Koz{\phi''}{\VV}.s}$ ~.
\item[]
\item\label{i1002}
    $\Koz{\CompNY{\mu}{X}{\phi}}{\VV} 
    \Wide{\Defs}
    \CompNY{\Lfp}{x}{\Koz{\phi}{\VV_{[X \mapsto x]}}}$
    \quad
    \parbox[t]{18em}{
     where by $\CompNY{\Lfp}{x}{\textit{exp}}$ we mean the least
     fixed-point of the function $\CompNY{\lambda}{x}{\textit{exp}}$.
    }
\item\label{i0943}
    $\Koz{\CompNY{\nu}{X}{\phi}}{\VV}
    \Wide{\Defs}
    \CompNY{\Gfp}{x}{\Koz{\phi}{\VV_{[X \mapsto x]}}}$ ~.
\end{enumerate}

Note that in the valuation $\VV_{[X \mapsto x]}$, the variable $X$ is mapped to the
expectation $x$.
\caption{\label{f1210} Kozen-style denotational semantics for \QTL}
\end{Figure}

\begin{lemma}\label{qTL well defined}
{\rm The quantitative logic \QTL\ is well-defined} --- \quad For any $\phi$ in the language,
and valuation $\VV$, the interpretation $\Koz{\phi}{\VV}$ is a well-defined expectation in $\E{S}$.

\begin{proof}\
Structural induction: arithmetic, that our formulae express only monotone functions, and that $(\E{S}$, $\PPle)$ is a
complete partial order. (Recall that $\E{S}$ is $[0,1]$-bounded.) 

\end{proof}
\end{lemma}

\section[Operational interpretation]{\begin{tabular}[t]{@{}l}
 Operational interpretation: \\
 \QMU\ generalises Stirling's game
\end{tabular}}
\label{game}\label{sec4}

In this section we give an alternative account of formulae $\phi$ (of the
reduced language), in terms of a generalisation of Stirling's turn-based game
\cite{LMCG}. The game is between two players, to whom we refer as
\MAX\ and \MIN. As in \Sec{qTL}, we assume a probabilistic transition system
$\PR{S}$ and a valuation $\VV$. Play progresses through a sequence of
\emph{game positions}, each of which is either a pair $(\phi, s)$ where $\phi$
is a formula and $s$ is a state in $S$, or a single $(y)$ for some real-valued payoff $y$
in $[0,1]$. Following Stirling, we will use the idea
of ``colours'' to handle repeated returns to a fixed point.

A sequence of game positions is called a \emph{game path} and is of the form $(\phi_0, s_0),~ (\phi_1, s_1),\ldots$ with (if finite) a payoff position $(p_n)$ at the end. The initial formula $\phi_0$ is the given $\phi$, and $s_0$ is an \emph{initial} state in $S$. A move from position $(\phi_i, s_i)$ to $(\phi_{i+1}, s_{i+1})$ or to $(y)$ is specified by the rules of \Fig{f0951}.

\begin{Figure}[p]

If the current game position is $(\phi_i, s_i)$, then play proceeds as follows:

\begin{enumerate}
\item\label{i1202} Free variables $X$ do not occur in the game --- their role is taken over by ``colours'' (see Cases \ref{i1740}--\ref{i1741a} below.).%
\footnote{Free variables do play a role in our more detailed analysis later (\Fig{f0856}).}
\item If $\phi_i$ is $\AAA$ then the game terminates in position $(y)$ where
$y = \VV.\AAA.s_i$.
\item[]
\item if $\phi_i$ is $\NextTime{\KK}\phi$ then the distribution $\VV.\KK.s_i$ is
used to choose either a next state $s'$ in $S$ or possibly the payoff state $\$$. If a state $s'$ is chosen, then the next game position is $(\phi, s')$; if $\$$ is chosen, then the next position is $(y)$, where $y$ is the payoff
$\VV.\KK.s.\$/(1-\sum_{s'\In S}\VV.\KK.s.s')$, and the game terminates.%
%
\item[]
\item If $\phi_i$ is $\phi' \Min \phi''$ (resp.~$\phi' \Max \phi''$) then \MIN\
(resp.~\MAX) chooses one of the minjuncts (maxjuncts): the next game position is
$(\phi, s_i)$, where $\phi$ is the chosen 'junct $\phi'$ or $\phi''$.
\item If $\phi_i$ is $\Bchoice{\GGG}{\phi'}{\phi''}$, the next game position is
$(\phi', s_i)$ if $\VV.\GGG.s_i$ holds, and otherwise it is $(\phi'', s_i)$.
\item[]
\item\label{i1740} If $\phi_i$ is $\CompNY{\mu}{X}{\phi}$ then a fresh colour $\CCC$ is chosen and is bound to the formula $\phi_{[X \mapsto \CCC]}$ for later use; the next game position is $(\CCC, s_i)$.%
\footnote{This use of colours is taken from Stirling
\cite{LMCG}; in \App{s1227a} we formalise the operations of choosing fresh colours and binding them to formulae.
The colour device easy determination, later on, of which recursion
operator actually ``caused'' an infinite path.}
\item\label{i1741} If $\phi_i$ is $\CompNY{\nu}{X}{\phi}$, then a fresh colour $\CCC$ is chosen and bound as for $\mu$.%
\footnote{The two kinds of fixed point are not distinguished at this stage: see \Def{d0920} below.}
\item[]
\item\label{i1741a} If $\phi_i$ is a colour $\CCC$, then the next game position is $(\phi, s_i)$, where $\Phi$ is the formula bound previously to $\CCC$.
\end{enumerate}

\bigskip
The game begins with a closed formula --- refer Item \ref{i1202}.~above.

Infinite games result in there being a single colour $\CCC$ that occurs infinitely often; finite games end in a payoff $(y)$ for $0 \leq y \leq 1$.

\caption{Rules for playing probabilistic formula-game.\label{f0951}}
\end{Figure}

A game path is said to be \emph{valid} if it can occur as a sequence according to
the above rules. Note that along any game path at most one colour can appear
infinitely often:
\begin{lemma}\label{one colour}
All valid game paths are either finite, terminating at some payoff $(y)$, or
infinite;
if infinite, then exactly one colour appears infinitely often.

\begin{proof}
Stirling \cite{LMCG}.
\end{proof}
\end{lemma}

To complete the description of the game, one would normally give the winning/losing conditions. Here however we are operating over real- rather than Boolean
values, and we speak of the ``value'' of the game. In the choices $\phi' \Max
\phi''$ (resp.~$\phi' \Min \phi''$) player \MAX\ (resp.~\MIN) follows a strategy in which he tries to maximise (minimise) a
real-valued ``payoff'' associated with the game,%
\footnote{In fact attributing the wins/losses to the two players makes it into
a \emph{zero-sum} game.}
defined as follows.
%
\begin{definition}\label{d0920} {\rm Value of a path} ---\quad 
The value of a path is determined by a fixed function $\Val$ defined by cases as follows:
\begin{enumerate}
\item The path $\pi$ is finite, terminating in a game state $(y)$; in this case the
value $\Val.\pi$ is $y$.

\item\label{i1056a} The path $\pi$ is infinite and there is a colour $\CCC$ appearing infinitely
often that was generated by a greatest fixed-point $\nu$; in this case $\Val.\pi$ is $1$.
\item\label{i1056b} The path $\pi$ is infinite and there is a colour $\CCC$ appearing infinitely
often that was generated by a least fixed-point $\mu$; in this case $\Val.\pi$ is $0$.
\end{enumerate}
\end{definition}

In \Sec{s1812} we make precise this notion of ``value of a game,'' and its interaction with strategies.

\section{Worked example: investing in the futures market}
\label{LSM}\label{sec5}

\subsection{Describing a game}

Typical properties of probabilistic systems are usually cost-based, and to illustrate that we give an example involving money. Concerning \emph{general} expected values, it lies strictly outside the scope of ``plain'' probabilistic temporal logic.

\begin{quotation}
\small
An investor $I$ has been given the right to make an investment in ``futures,'' a fixed number of shares in a specific company that he can \emph{reserve} on the first day of any month he chooses. Exactly one month later, the shares will be \emph{delivered} and will collectively have a market \emph{value} on that day --- he can \emph{sell} them then if he wishes.

His problem is to decide when to make his reservation so that the subsequent sale has maximum value.%
\footnote{\label{n1300}At ($+$) in \Sec{sec7} we discuss the related problem of maximising \emph{profit}.}
\end{quotation}

\noindent
The details are as follows:
\begin{enumerate}
\item\label{i1037a} The market value $v$ of the shares is a whole number of dollars between \$0 and \$10 inclusive; it has a probability $p$ of going up by \$1 in any month, and $1{-}p$ of going down by \$1 --- but it remains within those bounds. The probability $p$ represents short-term market uncertainty.
\item Probability $p$ itself varies month-by-month in steps of $0.1$ between zero and one: when $v$ is less than \$5 the probability that $p$ will rise is $2/3$; when $v$ is more than \$5 the probability of $p$'s falling is $2/3$; and when $v$ is \$5 exactly the probability is $1/2$ of going either way. The movement of $p$ represents investors' knowledge of long-term ``cyclic second-order'' trends. 
\item\label{i1044a} There is a cap $c$ on the value of $v$, initially \$10, which has probability $1/2$ of falling by \$1 in any month; otherwise it remains where it is. (This modifies Item \ref{i1037a}.\ above.) The ``falling cap'' models the fact that the company is in a slow decline.
\item If in a given month the investor does not reserve, then at the very next month he might find he is temporarily \emph{barred} from doing so. But he cannot be barred two months consecutively.
\item If he \emph{never} reserves, then he never sells and his return is thus zero.
\end{enumerate}

If it were not for Item \ref{i1044a}., the investor's strategy would be the obvious ``wait until $v\geq 9 \And p=1$ --- however long that takes ---  and make a reservation then.'' But the falling cap defeats that, effectively discounting the payoff as time passes.%
\footnote{\label{n0951}With the cap $c$ fixed at \$10 we know that from any state there is a non-zero probability, however small, of reaching $v\geq 9 \And p=1$ eventually; but, with the Zero-One Law \cite{ToPCP,PRfPL,ARaPfPS} for probabilistic processes, that means in fact that $v\geq 9 \And p=1$ will be reached eventually with probability one. So ``waiting'' would be the correct strategy, because when $v\geq 9 \And p=1$ finally occurs, an immediate reservation is guaranteed to pay \$10 in a month's time.}
Below we consider more sophisticated strategies that take that into account.

\bigskip
The situation is summed up by the transition system set out in \Fig{Commodity
futures}:
\begin{itemize}
\item During each month there are three purely probabilistic actions that occur, and their compounded effects determine a transition $m$, which we will call \textsf{month} in our formula (to come);
\item At the beginning of each month, the investor makes a maximising (angelic) choice of whether to reserve; but, if he does not, then 
\item At the beginning of the next month, there is a minimising (demonic) choice of whether he is barred.
\end{itemize}
\begin{Figure}[p]
\begin{center}
\epsfxsize=20em \epsfbox{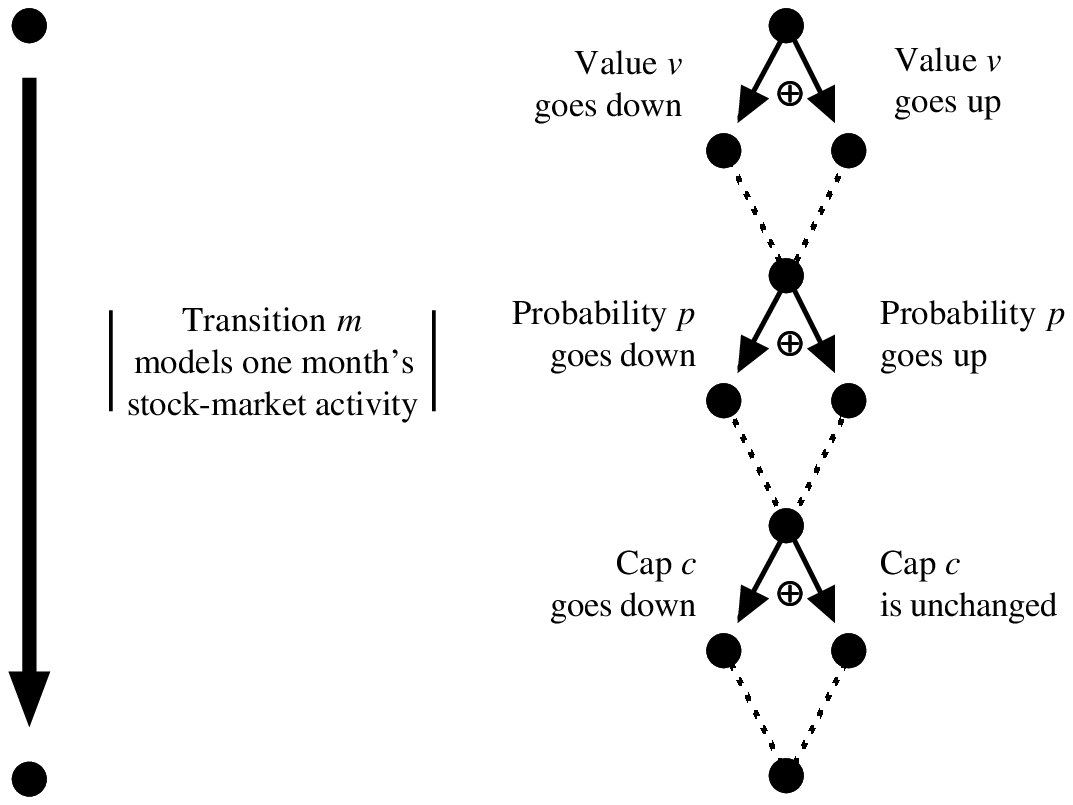}
\end{center}
The thick arrow, ``exploded'' on the right, represents the effect of one month's stock-market activity: symbol $\oplus$ labels the probabilistic choices it entails. The share value $v$ may rise or fall, according to $p$; probability $p$ itself may rise or fall, according to long-term trends; and the capped value $c$ of the stock may fall.

\bigskip
\begin{center}
\epsfxsize=30em \epsfbox{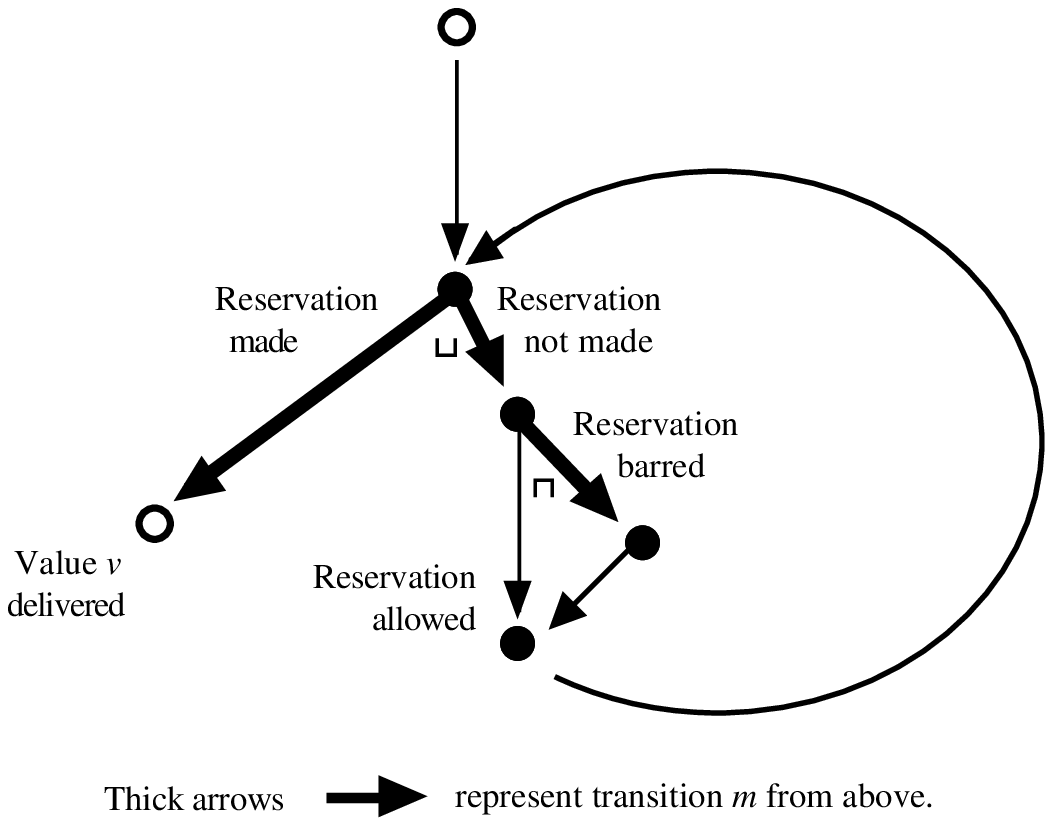}
\end{center}
This represents the non-deterministic choices available to the investor (maximising player) and the market (minimising player). Symbol $\Max$ represents the investor's choice; symbol $\Min$ represents the market's choice.

\bigskip
(The probabilistic choices occur ``within'' the thick arrows.)
\caption{Futures trading on the stock market: example.}
\label{Commodity futures}
\end{Figure}

The utility of our game interpretation in \Sec{game} is that we can easily use
the intuition it provides to write a formula describing the above system. The state space is $(v,p,c)$, and we use a transition
\newcommand{\Month}{\textsf{month}}
\newcommand{\Sold}{\textsf{Sold}}
\[
 \begin{array}{l@{\quad}c@{\quad}l}
  m & \Defs &
   \begin{array}[t]{@{}l}
    v \Gets (v+1) \Min c ~~\PC{p}~~ (v-1) \Max 0; \\\\
    \begin{array}{@{}l@{~~}c@{~~}l@{~~}l}
     \textbf{if} & v<5 & \textbf{then}
      & p\Gets (p{+}0.1) \Min 1 ~\PC{2/3}~ (p{-}0.1) \Max 0 \\
     \textbf{elsif} & v>5 & \textbf{then}
      & p\Gets (p{-}0.1) \Max 0~\PC{2/3}~ (p{+}0.1) \Min 1 \\
     \textbf{else} &&
      & p\Gets (p{-}0.1) \Max 0 ~\PC{1/2}~ (p{+}0.1) \Min 1 \\
     \textbf{fi};
    \end{array} \\\\
    c \Gets (c{-}1) \Max 0 ~~\PC{1/2}~~ c
   \end{array} 
 \end{array}
\]
to capture the effect of the large arrows.

\renewcommand{\Game}{\textit{Game}}
We can then use our (reduced) logical language
to describe the surrounding angelic and demonic choices, including the ``loop back'' (fixed point) which gives value zero (\IE\ $\mu$) if it never terminates. Using \Month\ to denote $m$, and a constant expectation \Sold%
\footnote{The \Sold\ function should have codomain $[0,1]$, but to avoid clutter
we have not scaled it down here by dividing by 10.} 
to denote the function $\widehat{v}$ returning just the $v$ component of the state, we would write our formula
as 
\begin{equation}\label{e0952}
 \Game \Wide{\Defs}
 \CompNY{\mu}{X}{\{\Month\}\Sold
                  ~\Max~
                 \{\Month\}(X \Min \{\Month\}X)} ~.
\end{equation}

\subsection{Playing the game}

Using the game interpretation, we can generate a probabilistic tree from the
transition system in \Fig{Commodity futures} by duplicating nodes with multiple incoming arcs (as in \Month), ``unfolding'' back-loops, and making minimising or maximising choices as they are encountered. In this example, at each unfolding both the investor $I$ (making $\Max$ choices) and the stock market $M$ (making $\Min$ choices whether to impose a bar) need to choose between two ongoing branches --- and their choice could be different each time
they revisit their respective decision points. Each of $I,M$ will be using a \emph{strategy}.

For example (recalling Footnote \ref{n0951}), the investor $I$'s  strategy (maximising, he hopes) for dealing with the falling cap might be
\begin{equation}\label{e1055}\parbox{0.8\textwidth}{
wait until the share value $v$ (rising) meets the cap $c$ (falling), and reserve then.
}\end{equation}
Waiting for $v$ to rise is a good idea, but when it has met the cap $c$ there is clearly no point in waiting further.

And $M$'s strategy (minimising, the investor fears) might be
\begin{equation}\label{e1055a}\parbox{0.8\textwidth}{
bar the investor, if possible, whenever the shares' probability $p$ of rising exceeds 1/2.
}\end{equation}

\bigskip
In general, let $\sigma_I$ and $\sigma_M$ be
sequences (possibly infinite) of choices, like the above, that $I$ and $M$ might make. When they
follow those sequences, the game-tree they generate determines a probability
distribution over valid game paths \cite{PaI}. Anticipating the next
section, let $\Tree{\phi}{\VV}{\sigma_I,\sigma_M}$ denote that path distribution%
\footnote{The distribution is described explicitly by \Def{d1127} and \Lem{l1127} still to come.}
as generated by $I$ and $M$'s choices. We can now describe $I$'s actual payoff as an expected value
\begin{equation}\label{e1416}
 Y.\sigma_I.\sigma_M
 \Wide{\Defs}
 \Exp{\Tree{\phi}{\VV}{\sigma_I,\sigma_M}}
    \textrm{``the $v$-component of the final state''} ~,
\end{equation}
with the understanding that the random variable in the integral's body yields zero
if in fact there is no final state (because of an infinite path).

In some cases, the choices made by $I$ and $M$ can be \emph{memoriless}, in the sense that in identical situations (identical values of $v,p,c$ in this case, and the same position in the transition system) they will always make the same choice. Both \Eqn{e1055} and \Eqn{e1055a} above are memoriless.%
\footnote{Adding \EG\ ``but reserve immediately if $v$ has fallen for five months in a row'' to Strategy \Eqn{e1055} would give it memory.}

Memoriless strategies are particularly
important for the efficient computation of expected payoffs \cite{CMDP}, and in
\Sec{dooalidy} we show that they suffice for analysis of \QTL\ formulae when the state space is finite --- that is, informally, the players gain no advantage by remembering where they have been.

\subsection{The value of the game}

As is usual in game theory, when the actual strategies of the two players are
unknown, we define the \emph{value} of the game to be the the
\emph{minimax} over all strategy sequences of the expected payoff --- but it is
well-defined only when it is the same as the \emph{maximin}, \IE\ when in the notation of \Eqn{e1416} we have
\[
 \Max_{\sigma_I}\Min_{\sigma_M} Y.\sigma_I.\sigma_M
 \Wide{=}
 \Min_{\sigma_M}\Max_{\sigma_I} Y.\sigma_I.\sigma_M ~.
\]

The equivalence proved in \Thm{t1416}, to come, tells us that such games' values are indeed well-defined, and that although we use the game interpretation to write down the formulae, we can use the logical interpretation
$\Koz{\Game}{\VV}$ to reason about their values. Sometimes, as in this simple case, we can use the logical interpretation to calculate an approximation directly.

\bigskip
Although the details of \Month\ are (deliberately) slightly messy, the structure of the overall formula \Game\ has been chosen (also deliberately) to be fairly simple,%
\footnote{It is in fact just a quantitative \emph{eventually} in the temporal subset \textit{qTL} of \QMU\ \cite{aEBMfPTL,aPTCboE}.}
and as such the fixed point can be approximated by iterating the function
\[
 \CompNY{\lambda}{X}{m\,\mbox{\scriptsize$\circ$}\,\widehat{v}
                     ~\Max~
                    m.(X \Min m.X)}
\]
beginning from the constant ``bottom'' function that is zero everywhere on the state space. 

Carrying out that calculation%
\footnote{We used \emph{Mathematica}$^\Registered$ for this example calculation, and the results were verified by Gethin Norman, at Birmingham University (UK), using the \emph{PRISM} model checker \cite{KNP02a} and MatLab$^\Registered$.
The scripts are available online \cite{qMuScripts}.}
shows for example that if $p$ is initially $0.5$ and the cap $c$ is 10, then the optimal expected sale-value for the investor is
\begin{equation}\label{e1455}
 \begin{array}{l|r|r|r|r|r|r|r|r|r|r|r|}
  \cline{2-12}
  \textit{\scriptsize initial share value:~~} &
   0&1&2&3&4&5&6&7&8&9&10 \\\cline{2-12}
  \textit{\scriptsize optimal expected sale:~~}
   & 4.16
   & 4.30
   & 4.55
   & 4.88
   & 5.24
   & 5.52
   & 6.00
   & 7.00
   & 8.00
   & 9.00
   & 9.50
  \\\cline{2-12}
 \end{array}
\end{equation}
Even when the share value is only ``moderately high,'' we thus see there is nothing to be gained by waiting, since the cap is likely to drop. For low initial values, however, some benefit can be gained by delaying the reservation for a while.

By comparison, the investor's ``seat-of-the-pants'' strategy at \Eqn{e1055} gives a significantly lower expected return against ``worst-case'' play by $M$:
\[
 \begin{array}{l|r|r|r|r|r|r|r|r|r|r|r|}
  \cline{2-12}
  \textit{\scriptsize initial share value:~~} &
   0&1&2&3&4&5&6&7&8&9&10 \\\cline{2-12}
  \textit{\scriptsize Strategy \Eqn{e1055}'s yield:~~}
   & 3.68
   & 3.79
   & 3.97
   & 4.17
   & 4.29
   & 4.17
   & 4.16
   & 4.65
   & 5.61
   & 6.78
   & 9.50
  \\\cline{2-12}
 \end{array}
\]
From this we might guess that when $v$ is at least \$6 (and $p,c$ are as given) it is better to ``reserve now'' (as \Eqn{e1455} suggests) than to follow Strategy \Eqn{e1055} and wait.

\subsection{Winning the game}\label{s0945}

Ideally we would like to be able to calculate both the value of the game and the strategies to realise it, \EG\ in this example we would like to be able to offer ``investment advice.'' Even better than knowing that Strategy \Eqn{e1055} can be improved, as we have just seen it can, is knowing \emph{how} to improve it.

In some cases, the logical interpretation can help by providing theorems that allow formulae to be simplified \cite{aEBMfPTL} or abstracted, thus bringing an apparently difficult formula within the range of probabilistic model-checkers \cite{PRISM}.

For formulae with a particularly simple structure, we might even be able to appeal to theorems --- proved in the logic --- which give maximising or minimising strategies directly. In the case of \Game, we do have such a theorem \cite{aPTCboE,ARaPfPS}: paraphrased, it states in this case that the investor should
\begin{equation}\label{e1034}\parbox{0.8\textwidth}{
make an immediate reservation just when the expected value of the stock in one month's time is at least as great as the expected value of the whole game played from this point.
}\end{equation}
Otherwise, he should wait.

The expected value of the stock in one month's time is easily calculated: it is just $m.\widehat{v}.(v,p,c)$, where $v,p,c$ are taken from the state ``now.'' (Note that the function $\widehat{v}$, as an argument of $m$, will instead take the $v$-value of the state in one month's time.) Tabulated for $p \Defs 0.5$ and $c \Defs 10$ as at \Eqn{e1455} above, that gives
\begin{equation}\label{e1455a}
 \begin{array}{l|r|r|r|r|r|r|r|r|r|r|r|}
  \cline{2-12}
  \textit{\scriptsize initial share value:~~} &
   0&1&2&3&4&5&6&7&8&9&10 \\\cline{2-12}
  \textit{\scriptsize
   \begin{tabular}{@{}l}
    expected share value \\
    in one month:
   \end{tabular}~~}
   & 0.50
   & 1.00
   & 2.00
   & 3.00
   & 4.00
   & 5.00
   & 6.00
   & 7.00
   & 8.00
   & 9.00
   & 9.50
  \\\cline{2-12}
 \end{array}
\end{equation}

Since the current values of $v,p,c$ are known at the beginning of each month (at the beginning of each turn, more generally), this maximising strategy can be applied in practice provided the fixed-point can be approximated sufficiently well. For our current game, comparing \Eqn{e1455} and \Eqn{e1455a} confirms our guess above about the problem with Strategy \Eqn{e1055}: instead of its recommendation, our initial move should be ``make an immediate reservation if $v \geq 6$, otherwise wait.''

In general, if we follow \Eqn{e1034} consistently we will realise at least the optimum \Eqn{e1455} over sufficiently many trials.

\subsection{Other games}\label{s1135}

Variations on \Eqn{e0952} can describe the value of other, related games.

It might be for example that a client's instructions are ``get me the shares when they're worth at least \$6,'' and the investor's aim is to maximise his chance of doing that.
Let \textsf{atLeast6} denote the characteristic function%
\footnote{Recall Footnote \ref{n1135}.}
of those states where $v\geq 6$; then
\[
 \CompNY{\mu}{X}{\{\Month\}\textsf{atLeast6}
                  ~\Max~
                 \{\Month\}(X \Min \{\Month\}X)} ~.
\]
gives a lower bound for $I$'s probability of achieving $v\geq6$ with an optimal strategy. By analogy with \Eqn{e1034} --- the same theorem applies --- that strategy should be 
\begin{quote}\parbox{0.8\textwidth}{
make an immediate reservation just when the probability of achieving $v\geq 6$ next month is at least as great as the optimal.
}\end{quote}

Below we tabulate the probabilities, giving for contrast the results of the strategy ``reserve when $v \geq 5$ and $p \geq 0.5$,'' \Ie\ the intuitive approach of waiting until the chance of achieving $v \geq 6$ next month is at least even:
\[
 \begin{array}{l|r|r|r|r|r|r|r|r|r|r|r|}
    \multicolumn{1}{}{~}
  & \multicolumn{11}{c}{\textit{\scriptsize probability of achieving $v \geq 6$}} 
  \\[0.5em]\cline{2-12}
  \textit{\scriptsize initial share value:~~} &
   0&1&2&3&4&5&6&7&8&9&10 \\\cline{2-12}
  \textit{\scriptsize
   \begin{tabular}{@{}l}
    following optimal strategy:
   \end{tabular}~~}
   & 0.25
   & 0.29
   & 0.34
   & 0.41
   & 0.46
   & 0.50
   & 0.56
   & 1.00
   & 1.00
   & 1.00
   & 1.00
  \\\cline{2-12}
  \textit{\scriptsize
   \begin{tabular}{@{}l}
    following intuitive strategy:
   \end{tabular}~~}
   & 0.25
   & 0.28
   & 0.33
   & 0.37
   & 0.42
   & 0.50
   & 0.50
   & 1.00
   & 1.00
   & 1.00
   & 1.00
  \\\cline{2-12}
 \end{array}
\]
We can see from the table that when $v$ is \$5 initially, the intuitive strategy is optimal: ``reserve now.'' At \$6, however, the optimal strategy --- counter-intuitively --- is to wait.

\subsection{More generally}

In the next section we show that the techniques used in this
example are valid for all games --- that is, that the value of any game of the form given
in \Sec{game} is well-defined, that it can be realised by memoriless
strategies if the state space is finite, and that its value corresponds exactly
to the denotational interpretation of \Sec{qTL}.

For the current example, those results justified our using the denotational
interpretation to analyse \Game, which in this
simple case led to a direct calculation \Eqn{e1455} of the optimal result, and the formulation of an explicit strategy \Eqn{e1034} to achieve it. For more complex formulae, the optimal
payoff can be determined indirectly using model-checking methods derived from Markov
Decision Processes (\textit{MDP}'s) \cite{CMDP}.

For example,
\emph{PRISM} \cite{KNP02a} is a probabilistic model checker which has support for \textit{MDP}'s.%
\footnote{It also handles discrete- and continuous-time Markov chains.}
It takes as input an \textsf{occam}-like \cite{occam} description of a transition system,
including both overlapping-guard style (traditional, in \textit{CSP} \cite{tTaPoC} parlance
``internal'') nondeterminism and (beyond \textsf{occam}/\textit{CSP}) probabilistic choice
constructs. Using \textit{BDD}-based techniques it translates the input
description into an \textit{MDP}, called $f$ say.

Normally, the tool allows the verification of \textit{MDP}'s against specifications written in
the temporal logic \textit{pCTL} \cite{ALRaTP}; in this case an extended version was used
that supports reward-based specifications.
The rewards are evaluated by approximating the least fixed-point of
$f_\sqcap$ or $f_\sqcup$ by repeated applications beginning from bottom (zero),
where $f_\sqcap$ or $f_\sqcup$ interprets (all) non-probabilistic nondeterminism as minimising, maximising respectively --- \Ie\ ``uni-modally'' --- and
the interpretation of the result as a measure of the minimum or maximum
possible reward in a probabilistic/demonic or probabilistic/angelic game
is justified by Everett's original work \cite{RG}.

\bigskip
To deal with the ``bi-modal,'' \emph{minimax} non-determinism of our example, the \emph{PRISM}-produced transition matrices for the \textit{MDP} were exported, and used as data for a MatLab$^\Registered$ program that performed the angelic/demonic calculations explicitly; the results
agreed with the calculations we had previously obtained from Mathematica$^\Registered$ by coding up the \QMU\ formula \Eqn{e0952} directly \cite{qMuScripts}.

The justification in this more general case that the value can be interpreted as the \emph{minimax} expected reward of the original game is provided by our \Thm{t1416} below, extending Everett.

\section{Proof of equivalence}\label{dooalidy}

In this section we give our main result, the equivalence of the operational, ``Stirling-game''
and the denotational, ``Kozen-logic'' interpretations of \QMU\ formulae. We formalise \emph{strategies} in both
cases,
whether they
can or cannot have ``memory'' of where the game or transition system has gone so
far, and the effect of ``\emph{minimaxing}'' over them.

To begin with, we fix a single pair of strategies: one maximising, one minimising.

\subsection{Fixed strategies for the Stirling interpretation} \label{s1812}

Our first step will be to explain
how the games of \Fig{f0951} can be formalised provided a fixed pair of players' strategies is decided
beforehand.

The current position of a game --- as we saw in \Sec{sec4} --- is a formula/state pair. We introduce two \emph{strategy functions} called $\Mins$ and $\Maxs$, which will prescribe \emph{in advance} the players' decisions to be taken as they go along: the functions are of type ``finite-game-path to Boolean,'' and the player \MIN\ (resp.\ \MAX), instead of deciding ``on the fly'' how to interpret a decision point $\Min$ (resp.\ $\Max$), takes the strategy function $\Mins$ (resp.\ $\Maxs$) and applies that to the sequence of game positions traversed so far. The result ``true'' means ``take the left subformula,'' say.

These strategies model full memory, because each is given as an argument the
complete history of the game up to its point of use. (Note that the history includes the current state $s$.) We stipulate however that strategies are \emph{colour-insensitive}, since colours are not an artefact of the system itself:%
\footnote{That is, since colours do not occur in the physical systems we are specifying, we are not obliged to model strategies that take them into account.}
that is, we assume that from any colour $\CCC$ it is possible to recover the identity of the variable $X$ for which it was generated, and any strategy treats game position $(\CCC,s)$ in a history as if it were $(X,s)$.%
\footnote{Thus strategies do not depend on the actual colour value that was arbitrarily chosen during a fixed-point step. All that matters is whether colours are the same or differ, and which kind of fixed point (least or greatest) generated them.}

\bigskip
We can now formalise our probabilistic extension of Stirling's game. Rather than see it as at our earlier \Fig{f0951}, a linear sequence of moves interleaving maximising, minimising and probabilistic choices, we use our strategy functions to present the game in two separated stages.

In the first stage we construct a (possibly infinite) purely probabilistic game-tree, using the given formula $\phi$, the initial
state $s$ and the pre-packaged strategy functions $\Mins,\Maxs$. The process is shown in \Fig{f0856}, and clearly is derived from the game of \Fig{f0951} given earlier: the difference is that in \Fig{f0856} the probabilistic choices are ``deferred'' by our showing the whole tree of their possibilities, whereas in \Fig{f0951} they are ``taken as they come.'' We write $\Tree{\phi}{\VV}{\Mins,\Maxs}.s$ for the tree generated by the process of \Fig{f0856}.

\begin{Figure}[p]
As for \Fig{f0951}, the current game position is some $(\phi,s)$; but here we appeal to pre-determined \emph{strategy} functions
$\Mins,\Maxs$, and use a ``current path'' variable $\pi$, to construct the whole probabilistic tree of possibilities rather than to play along one of its branches as we go.

\bigskip
After each step, the path $\pi$ is extended with $(\phi,s)$; it is initially empty. The formula and state change as for \Fig{f0951}, but according to the given strategies 
if appropriate.

\begin{enumerate}
\item\label{i1250} If $\phi$ is a free variable $X$, make a single probability-one edge leading to tip $(\VV.X.\pi.s)$.%
%
\item\label{i1827} If $\phi$ is $\AAA$ then make a single probability-1 edge leading to tip $(\VV.\AAA.s)$.
\item[]
\item if $\phi$ is $\NextTime{\KK}\Phi$ then make one edge for each state $s'$ having  $\VV.\KK.s.s'$ non-zero, labelling it with that probability, plus one more ``payoff'' edge if those probabilities sum to less than one. For each $s'$ edge, add a child $(\Phi, s')$; if there is a payoff edge then add a child $(y)$ where $y$ is the payoff $\VV.\KK.s.\$/(1-\sum_{s'\In S}\VV.\KK.s.s')$.%
\footnote{If the probabilities over $S$ sum to one then we do not add a payoff edge --- so the question of division by zero does not arise.}

%
\item[]
\item If $\phi$ is $\Phi' \Min \Phi''$ (resp.\ $\Phi' \Max \Phi''$) then choose between $\Phi'$ and $\Phi''$ depending on $\Mins.\pi$ (resp. $\Maxs.\pi$): form a single edge of probability one to the next game position $(\Phi, s)$, where $\Phi$ is the chosen 'junct $\Phi'$ or $\Phi''$.
\item If $\phi$ is $\Bchoice{\GGG}{\Phi'}{\Phi''}$, choose between $\Phi'$ and $\Phi''$ depending on $\VV.\GGG.s$: form a single edge of probability one to the next game position $(\Phi, s)$, where $\Phi$ is the chosen 'junct.
\item[]
\item If $\phi$ is $\CompNY{\mu}{X}{\Phi}$ then choose fresh
colour $\CCC$; make a single probability-one edge leading to $(\CCC, s)$.
\item If $\phi$ is $\CompNY{\nu}{X}{\Phi}$ then (as for $\mu$) choose fresh
colour $\CCC$; make a single probability-one edge leading to $(\CCC, s)$.
\item[]
\item\label{i1107} If $\phi$ is colour $\CCC$, extract the game position $(\CompNY{\mu/\nu}{X}{\Phi},s')$; make a single probability-one edge leading to $(\Phi[X \mapsto \CCC], s)$.
\end{enumerate}

\bigskip
\App{s1227a} at ($+$) explains how the operations of choosing and binding colours are formalised in this denotational definition.

\bigskip
We write $\Tree{\phi}{\VV}{\Mins,\Maxs}.s$ for the tree generated, as above, from formula $\phi$, strategies $\Mins,\Maxs$ and initial state $s$.

\caption{Tree-building process, with paths and strategies.\label{f0856}}
\end{Figure}

For the second stage we play the purely-probabilistic game represented by the tree just generated, and use the function $\Val$ of \Def{d0920}, from valid game paths to
the non-negative reals, to determine the ``payoff'' as described at the end of \Sec{sec4}. Our ``expected payoff'' from the whole process is then the expected value of this payoff function over the distribution of paths determined by that game tree, formalised as follows. (Abusing notation, we write $\Tree{\phi}{\VV}{\Mins,\Maxs}.s$ for the probability distribution of paths determined by the tree, as well as for the tree itself.)

\begin{definition}\label{d1127} {\rm Value of fixed-strategy Stirling game} ---\quad The value of a game played from formula $\phi$ and initial state $s$, with fixed strategies $\Mins,\Maxs$, is given by the expected value
\[
 \Exp{\Tree{\phi}{\VV}{\Mins,\Maxs}.s} \Val
\]
of $\Val$ over the (probability distribution determined by the) game-tree $\Tree{\phi}{\VV}{\Mins,\Maxs}.s$ generated by the formula, the strategies and the initial state as shown in \Fig{f0856}.

(The argument that this is well defined is given in \Lem{l1127} following.)
\end{definition}

\begin{lemma}\label{l1127} {\rm Well-definedness of \Def{d1127}} ---\quad
The expected value of $\Val$ over game trees is well-defined.

\begin{proof}
We must show (1) that $\Tree{\phi}{\VV}{\Mins,\Maxs}.s$ generated as at \Fig{f0856} determines a sigma-algebra, and (2) that $\Val$ is measurable over it. We use in several places that the tree is finitely branching, and that therefore it has only countably many nodes (and hence only countably many finite paths).

\bigskip
For (1) we appeal to the standard construction of path distributions from trees: the basis elements are ``cones'' of paths all having a common (finite) prefix; and the measure of a cone is the product of the probabilities found on the path leading from the root to the end of the common prefix (equivalently, to the base of the cone). The algebra is generated by closing the basis under countable unions and complement (and hence countable intersections also).

\bigskip
For (2) we must show that for any real $r$ the inverse image $\Val^{-1}.(r,\infty)$ is in the algebra defined at (1), where $(r,\infty)$ is the open interval of reals above $r$.

We begin with the case $0<r<1$, in which case the inverse image is the set of all paths containing an infinite number of $\nu$-colours \emph{plus} all the (finite) paths ending in an explicit $s$-tip with $r<s$. But since there are only countably many finite paths in total, there are certainly only countably many $(s)$-tipped paths --- which we can therefore ignore.

Since each new colour (of either kind) is generated at some node of the tree, there are only countably many $\nu$-colours, and so we may concentrate on a single $\nu$-colour $\CCC$.

For any $i \geq 0$ the set $\CCC_i$ of paths with at least $i$ occurrences of $\CCC$ is measurable, since it is the union of all cones determined by finite prefixes ending in an $i^{\it th}$ occurrence of $\CCC$ exactly. Then the set $\CCC_\infty$ of paths containing infinitely many $\CCC$'s is just the countable-over-$i$ intersection of all the $\CCC_i$'s.

\bigskip
We finish by noting that for the case $1\leq r$ the inverse image is empty; and for the case $r=0$ it is just the set of all paths.
\end{proof}
\end{lemma}

\bigskip
Although the game is played ``all at once'' in \Fig{f0951}, note that the strategy functions and the construction of \Fig{f0856} make it appear as if it is played in two stages: first, we determine the strategies; second, we roll the dice. The point of that is to allow us to use standard techniques of expected values in the second, purely probabilistic stage, free of the complications of max/min-nondeterminism. The strategy functions' generality makes the two views equivalent.

\subsection{Fixed strategies for the Kozen interpretation}

Now that the value of a fixed-strategy game is defined,
our second step is to define fixed-strategy denotations: we augment the semantics of
\Sec{sec3} with the same strategy functions as above. For clarity we use slightly
different brackets $\KozS{\phi}{\VV}{\Mins, \Maxs}$ for the extended semantics.

The necessary alterations to the rules in \Fig{f1210} are straightforward, the
principal one being that in Case \ref{i1219}, instead of taking a minimum or
maximum, we use the argument $\Mins$ or $\Maxs$ as appropriate to determine
whether to carry on with $\phi'$ or with $\phi''$.

A technical complication is
then that all the definitions have to be changed so that the ``game sequence so
far'' is available to $\Mins$ and $\Maxs$ when required. That can be arranged
for example by introducing an extra ``path-so-far'' argument and passing it, suitably extended, on every right-hand side. 

The modified rules are given in full in \App{s1227a} at \Fig{f0949}.

\subsection{Equivalence of the interpretations}

We now have our first equivalence, for fixed strategies:
\begin{lemma}\label{l1227} {\rm Equivalence of fixed-strategy games and logic} ---\quad
For all closed \QTL\ formulae $\phi$, valuations $\VV$, states $s$ and strategies
$\Mins,\Maxs$, we have
\[
 \Exp{\Tree{\phi}{\VV}{\Mins,\Maxs}.s} \Val
 \quad=\quad
 \KozS{\phi}{\VV}{\Mins,\Maxs}.s ~.
\]

\begin{proof} (sketch%
\footnote{A full proof is given in \App{s1227a}.})
The proof is by structural induction over $\phi$, straightforward
except when least- or greatest fixed-points generate infinite trees. In those
cases we consider approximations  to the valuation function $\Val$ such that $\Val^{\,\CCC}_n.\pi$ acts as $\Val$ if path $\pi$ contains less than $n$ occurrences of colour $\CCC$, otherwise returning zero (resp.~one) for the $\mu$ (resp.~$\nu$) cases respectively. Those $n$-approximants in the game interpretation are shown by mathematical induction to correspond to the usual $n$-fold iterates that approximate fixed points in the denotational interpretation; and bounded monotone convergence \cite{PaI} is used to distribute suprema (for least fixed-points) through $\Exp{}$.

For similar distribution of the infima required by greatest fixed-points, we subtract from one and again argue over suprema.
\end{proof}
\end{lemma}

\Lem{l1227} will be the key to our completing the argument --- in \Sec{s1234} to follow --- that the value of the Stirling
game is well-defined when we take the \emph{minimax}
over \emph{all} strategies of the expected payoff, rather than just considering a fixed pair. That is, in the notation of this section we must establish
\begin{equation}\label{e2208}
 \Min_\Mins \Max_ \Maxs \Exp{\Tree{\phi}{\VV}{\Mins,\Maxs}.s} \Val
 \Wide{=}
 \Max_\Maxs \Min_ \Mins \Exp{\Tree{\phi}{\VV}{\Mins,\Maxs}.s} \Val ~.
\end{equation}
The utility of \Lem{l1227} is that it allows us to carry out the argument in a
denotational rather than operational context --- we can avoid the integrals, games and trees and simply use $\KozS{\cdot}{}{}$ and \emph{cpo}'s instead.

In fact we show \Eqn{e2208} to be even simpler --- both sides are equal to the original 
denotational interpretation,
with its $\Min$ and $\Max$ operators still in the formula and therefore no need for strategy functions at all. That is, we prove \Eqn{e2208} by appealing to \Lem{l1227} to move from $\int$'s to $\KozS{\cdot}{}{}$'s, and then we will establish the equality
\[
 \Min_\Mins \Max_ \Maxs \KozS{\phi}{\VV}{\Mins,\Maxs}
 \Wide{=}
 \Koz{\phi}{\VV}
 \Wide{=}
 \Max_\Maxs \Min_ \Mins \KozS{\phi}{\VV}{\Mins,\Maxs}
 ~.
\]
And so we will have that the game is indeed well-defined --- and that $\Koz{\phi}{\VV}$ is its value.

\subsection[Full equivalence \emph{via} memoriless strategies]{\begin{tabular}[t]{@{}l}
Full equivalence \emph{via}  memoriless strategies \\
for finite state spaces
\end{tabular}}\label{s1234}

In the previous section we handled maximising/minimising strategies by modelling them explicitly as a fixed pair of ``decision'' functions chosen beforehand. Here we show that the order in which they are chosen makes no difference: whether \emph{max}-before-\emph{min} or the reverse, the Stirling-value of the game equals the Kozen-value of the original formula, \IE\ with the $\Max/\Min$ operators still in place and no explicit strategy functions.

A key step in that process is showing that, over a finite state space, there are fixed ``memoriless'' strategies that ``solve'' a Kozen interpretation in the sense of achieving its value by local decisions that depend only on the current state and not on the history; such strategies are implemented by Boolean conditionals.

Our approach is a generalisation of an argument used by Everett, who treated formulae with a single least fixed-point \cite{RG}; we have generalised it to deal with multiple fixed points nested arbitrarily.

\bigskip
Let formula $\phi_{\UnderB}$ be derived from $\phi$ by the syntactic operation of replacing each
operator $\Min$ in $\phi$ by a specific predicate symbol drawn from a tuple
$\UnderB$ of our choice, possibly a
different symbol for each syntactic occurrence of $\Min$. This represents replacing the general minimising strategy $\Min$ by some specific memoriless strategy (-ies) $\underline{G}$ that $\UnderB$ denotes.

Similarly we write $\phi_{\OverB}$
for the derived formula in which all instances of $\Max$ are replaced left-to-right by 
successive predicate symbols in a tuple $\OverB$.

With those conventions,
we will appeal to \Lem{Everett} of \App{s1650} that for
all \QTL\ formulae $\phi$ over a finite%
\footnote{Finiteness is needed in Case $\OverB$ of the lemma's proof.}
state space $S$,
and valuations $\VV$, there exist (semantic) predicate tuples $\underline{G}$
and $\overline{G}$ corresponding to the predicate symbols as above such that
$
   \Koz{\phi_{\UnderB}}{\VV'} 
   =  
   \Koz{\phi}{\VV} 
   = 
   \Koz{\phi_{\OverB}}{\VV'}
$~,
where $\VV'$ is the technical extension of $\VV$ that maps the new symbols $\UnderB,\OverB$ to $\underline{G},\overline{G}$ respectively, and leaves all else unchanged.

\bigskip
For example, if the formula $\phi$ is 
\[
 \CompNY{\mu}{X}{\AAA_1 \Max
 \CompNY{\nu}{Y}{\AAA_2 \Min \{\KK\}(\AAA_3 \Max (\Bchoice{\GGG}{X}{Y}))}}~,
\]
then we are saying we can find
predicate tuples
$(\underline{G}_1)$ and $(\overline{G}_1, \overline{G}_2)$
so that for corresponding predicate-symbol tuples
$\UnderB \Defs (\UnderB_1)$ and $\OverB \Defs (\OverB_1, \OverB_2)$ we can define
\[
 \begin{array}{lcll}
  \phi_{\UnderB} & \Wide{\Defs}
   & \CompNY{\mu}{X}{\AAA_1 \Max      \CompNY{\nu}{Y}{\Bchoice{\UnderB_1}{\AAA_2}{\{\KK\}(\AAA_3 \Max (\Bchoice{\GGG}{X}{Y}))}}}
   & \textrm{\quad and} \\
   \phi_{\OverB} & \Wide{\Defs}
   & \CompNY{\mu}{X}{\Bchoice{\OverB_1}{\AAA_1} {\CompNY{\nu}{Y}{\AAA_2 \Min \{\KK\}(\Bchoice{\OverB_2}{\AAA_3} {(\Bchoice{\GGG}{X}{Y})})}}}
 \end{array}
\]
--- and then extend $\VV$ to a $\VV'$ that
takes $\UnderB_1, \OverB_1, \OverB_2$ to $\underline{G}_1, \overline{G}_1, \overline{G}_2$ respectively ---
so that $\phi_{\UnderB}$, $\phi_{\OverB}$
and $\phi$ itself are all $\Koz{\cdot}{\VV'}$-equivalent. \footnote{It is easy to show also that all three formulae are then $\Koz{\cdot}{\VV'}$-equivalent to $\phi_{\UnderB,\OverB}$~, but we do not need that.}

The proof of \Lem{Everett} is by induction, intricate only in one case, which is where we rely on
Everett's techniques [\textit{op.~cit}].%
\footnote{Unfortunately Everett's work as it stands is less than we
need, so although we borrow his techniques we cannot simply appeal to his result as a whole.}
That part of the proof, together with several preliminary lemmas, is given in Appendices \ref{s1650}--\ref{a1250}.

With it we show, first, that the Kozen interpretation is insensitive to the order in which the strategies are chosen; then, from \Lem{l1227} we have immediately that Stirling games are similarly insensitive --- and thus our main result, that the value of the game is the value of the denotation.

\begin{lemma}\label{l1322} {\rm \emph{Minimax} equals \emph{maximin} for Kozen interpretation} ---\quad

For all \QTL\ formulae $\phi$, valuations $\VV$ and strategies $\Mins,\Maxs$, we
have 
\begin{equation}\label{e1149}
  \Min_\Mins \Max_ \Maxs \KozS{\phi}{\VV}{\Mins,\Maxs} \\
  \Wide{=}
  \Max_\Maxs \Min_ \Mins \KozS{\phi}{\VV}{\Mins,\Maxs}~.
\end{equation}

\begin{proof}
From monotonicity, we need only prove $\LHS \leq \RHS$.%
\footnote{Trivially \(
\Min_\Mins \Max_ \Maxs \KozS{\phi}{\VV}{\Mins,\Maxs}
~=~
\Max_ \Maxs \Min_\Mins \Max_ \Maxs \KozS{\phi}{\VV}{\Mins,\Maxs}
~\geq~
\Max_ \Maxs \Min_\Mins \KozS{\phi}{\VV}{\Mins,\Maxs}~.
\)}
Note that from
\Lem{Everett} we have predicates $\overline{G}$
and $\underline{G}$ satisfying
\begin{equation}\label{everett1322}
   \Koz{\phi_{\UnderB}}{\VV'} 
   \Wide{=}  
   \Koz{\phi}{\VV'} 
   \Wide{=} 
   \Koz{\phi_{\OverB}}{\VV'} ~,
\end{equation}
with $\VV'$ extending $\VV$ as we have said, a fact which we use further below.

To begin with, using the predicates $\UnderB$ from \Eqn{everett1322}, we start from the \LHS\ of \Eqn{e1149} and observe that
\begin{equation}\label{e1354}
 \Min_\Mins \Max_ \Maxs \KozS{\phi}{\VV}{\Mins,\Maxs} \\
 \Wide{=}
 \Min_\Mins \Max_ \Maxs \KozS{\phi}{\VV'}{\Mins,\Maxs} \\
 \Wide{\leq}
 \Max_\Maxs \KozS{\phi_{\UnderB}}{\VV'}{\Maxs}~,
\end{equation}
--- in which on the right we omit the now-ignored $\Mins$ argument --- because (on the left) formula $\phi$ does not refer to the extra symbols in $\VV'$ and (on the right) the $\Min_\Mins$ can
select exactly
those predicates $\underline{G}$ referred to in $\VV'$ by $\UnderB$ simply by making an appropriate choice of
$\Mins$.

We then eliminate the explicit strategies altogether by observing that
\begin{equation}\label{e1355}
 \Max_\Maxs \KozS{\phi_{\UnderB}}{\VV'}{\Maxs}
 \Wide{\leq}
 \Koz{\phi_{\UnderB}}{\VV'}~,
\end{equation}
because the simpler $\Koz{~}{}$-style semantics on the right interprets $\Max$
as maximum, which cannot be less than the result of appealing to some strategy
function $\overline{\sigma}$.

We can now continue on our way towards the \RHS\ of \Eqn{e1149} as follows:
\begin{Reason}
\StepR{}{carrying on from \Eqn{e1355}}{
    \Koz{\phi_{\UnderB}}{\VV'}
}
\StepR{$=$}{first equality at \Eqn{everett1322}}{
    \Koz{\phi}{\VV'}
}
\StepR{$=$}{second equality at \Eqn{everett1322}}{
    \Koz{\phi_{\OverB}}{\VV'}
}
\StepR{$\leq$}{as for \Eqn{e1355} above, backwards and with inequality reversed}{
    \Min_\Mins \KozS{\phi_{\OverB}}{\VV'}{\Mins}
}
\StepR{$\leq$}{as for \Eqn{e1354} above, backwards and with inequality reversed}{
    \Max_\Maxs \Min_\Mins \KozS{\phi}{\VV}{\Mins,\Maxs} ~,
}
\end{Reason}
and we are done. (Note that in the last step we were again able to use the fact that $\phi$ is insensitive to the difference between the extended valuation $\VV'$ and the original valuation $\VV$.)
\end{proof}
\end{lemma}

The proof above establishes the duality we seek between the two interpretations, and our principal result:

\begin{theorem}\label{t1416} {\rm Value of Stirling game} --- \quad
The value of a Stirling game is well-defined, and equals
$\Koz{\phi}{\VV}$.

\begin{proof}
\Lem{l1227} and \Lem{l1322} establish the equality \Eqn{e2208}, for well-definedness; the stated equality with $\Koz{\phi}{\VV}$
occurs
during the proof of the latter.
\end{proof}
\end{theorem}

Finally, we have an even tighter result about the players' strategies:

\begin{lemma}\label{memoriless} {\rm Memoriless strategies} --- \quad
There exists a memoriless strategy $\overline{G}$ which, if followed by player \MAX,
achieves the value of the Stirling game against all strategies of player \MIN. (A
similar result
holds for player \MIN.)

\begin{proof}
Directly from \Lem{Everett} and \Thm{t1416}.
\end{proof}
\end{lemma}

\section{Conclusion}
\label{sec7}

Von Neumann and Morgenstern \cite{ToGaEB} proved the \emph{minimax} theorem for zero-sum two-player games comprising one kind of play in a single game. Everett \cite{RG} extended this to ``least-fixed point'' games, \Ie\ an unbounded number of plays of a finite number of possibly different games that can recursively call each other within a single ``loop.'' For the special case where those games are turn-based, we have extended that result further to include both least- and greatest fixed points, and arbitrary nesting.

Our reason for doing this was to introduce a novel game-based interpretation for the quantitative $\mu$-calculus \QTL\ over probabilistic/angelic/demonic transition systems, probabilistically generalising Stirling's
game interpretation of the standard $\mu$-calculus; we aimed to show it equivalent to our existing Kozen-style interpretation of \QTL, and so to provide an ``operational'' semantics.

The equivalent interpretations are general enough to specify cost-based properties of
probabilistic systems --- and many such properties lie outside
standard temporal logic. The Stirling-style interpretation is close to automata-based
approaches, whilst the Kozen-style logic (studied more extensively elsewhere
\cite{aEBMfPTL}) provides an attractive proof system.

Part of our generalisation has been to introduce the Everett-style ``payoff states'' $\$$ into Stirling's generalised games. Although many presentations of probabilistic transitions (including our earlier work) do not include the extra state, giving instead simply functions from $S$ to $\Lift{S}$ which in effect take the primitive elements of formulae to be probabilistic programs, here our primitive elements are small probabilistic \emph{games} \cite{RG}. The probabilistic programs are just the simpler special case of payoff zero. The full proof \cite{proofs} of \Lem{Everett} makes that necessary, since we treat the $\UnderB$/$\nu$ case \emph{via} a duality, appealing to the $\OverB$/$\mu$ case. But it is a duality under which probabilistic programs are not closed, whereas the slightly more general probabilistic games \emph{are} closed. Thus we have had to prove a slightly more general result.

\bigskip
An interesting possibility for further work is the use of \emph{intermediate} fixed points, yielding say a value $0<x<1$ rather than the fixed zero-for-least and one-for-greatest that are traditional. For expectation transformer $t$ we would propose the definition
\begin{equation}\label{e0931}
 \textsf{fix}_x.t
 \Wide{\Defs}
 \lim_{n\rightarrow\infty} t^n.\Const{x} ~,
\end{equation}
where $\Const{x}.s \Defs x$,
so that (for continuous $t$ at least) $\mu.t,\nu.t$ become the special cases of zero and one for $x$, \IE\ $\textsf{fix}_0.t$ and $\textsf{fix}_1.t$. When $t$ is purely probabilistic (thus almost linear), it can be shown that \Eqn{e0931} is meaningful (\IE\ converges) for any $0 \leq x \leq 1$, and agrees with $\mu,\nu$ where it should.%
\footnote{Using a constant expectation $\Const{x}$ is necessary, as $\lim_{n\rightarrow\infty} t^n.e$ does not converge in general if expectation $e$ may vary over the state.

For example let $S \Defs \{0,1\}$ and take $t$ to be (the transformer corresponding to) $s\Gets 1{-}s$ for $s \in S$, with $e.s \Defs s$; then $t^n.e.s = (s+n) \mathbin{\textbf{mod}} 2$.}  

We do not know however whether convergence is guaranteed when $t$ may contain angelic or demonic nondeterminism.

\bigskip
The utility of $\textsf{fix}_x$ is when infinite behaviour is to attract a reward which is neither zero nor one. In the game interpretation we would collapse Cases \ref{i1056a},\ref{i1056b} of \Def{d0920} to the single
\begin{itemize}
\item[\ref{i1056a}.] The path $\pi$ is infinite and there is a colour $\CCC$ appearing infinitely
often that was generated by $\textsf{fix}_x$ for some $x$; in this case $\Val.\pi$ is $x$.
\end{itemize}
In the logical interpretation we would use \Eqn{e0931} just above.

\noindent\makebox[0pt][r]{$+$\quad}\hspace{\parindent}%
For the investor of \Sec{LSM} it might be that his reservation costs some fixed \$$x$, so that infinite behaviour (never reserving) is awarded \$$x$ rather than zero (\IE\ he keeps his money). A more advanced use would be that he seeks to maximise his \emph{profit},%
\footnote{Recall Footnote \ref{n1300}.}
defined to be the difference $v_1 - v_0$, where $v_0$ is the market value $v$ when he reserves, and $v_1$ is its value one month later (when the shares are delivered, and he can sell). Because $v_1-v_0$ could be negative, we would shift-and-scale to transform the expectations into the range $[0,1]$, with the effect that the zero awarded for ``never reserves'' would be transformed to $0.5$.
%


\section{Related work}

Probabilistic temporal logics, interpreted over nondeterministic/probabilistic
transition systems, have been studied extensively, most notably by de Alfaro
\cite{TLftSoPaR}, Jonsson \cite{aLfRaTaR}, Segala \cite{MaVoRDRTS} and Vardi
\cite{aTFC}. Condon \cite{tCoSG} considered the complexity of underlying transition systems like ours, including probabilistic- (but $\PC{1/2}$ only), demonic- and angelic choice, but without our more general expectations and payoffs. Monniaux \cite{AIoPS} uses Kozen's deterministic formulation together with demonic program inputs to analyse systems \emph{via} abstract interpretation \cite{AIF}.

The \emph{pCTL} of Aziz \cite{IUW} and Hansson and Jonsson \cite{ALRaTP} provides a threshold operator which allows properties such as ``$\phi$ is eventually satisfied with probability at least 0.75,'' where the underlying distribution is over execution paths.
Similarly Narasimha \EtAl\
\cite{PTLvtMMC} use probability thresholds, and restrict to the alternation-free
fragment of the $\mu$-calculus; for that fragment they do provide an
operational interpretation which selects the proportion of paths that satisfy
the given formula. Their transition systems are deterministic.

Though the quantitative $\mu$-calculus has received much less attention, its use of expected values allows a greater variety of expression --- in particular, it can specify properties that are inherently cost-based.

Huth and Kwiatkowska \cite{QAaMC} for example use real-valued expressions based on
expectations, and they have investigated model-checking approaches to evaluating
them; but they do not provide an operational interpretation of the logic, nor
have they exploited its algebraic properties \cite{aEBMfPTL}. 

\bigskip
De Alfaro and Majumdar \cite{QSoORG} use \QTL\ to address an issue similar to,
but not the same as ours: in the more general context of concurrent games, they
show that for every LTL formula $\Psi$ one can construct a \QTL\ formula $\phi$
such that $\Koz{\phi}{\VV}$ is the greatest assured probability that Player 1
can force the game path to satisfy $\Psi$.

\newcommand{\AtB}{\textsf{atB}}
The difference between de Alfaro's approach and ours can be seen by considering the formula $\Psi \Defs
\CompNY{\mu}{X}{\NextTime{\KK}\AtB \Max \NextTime{\KK}X}$ over the transition
system
\[
 \VV.\KK \Wide{\Defs} \Cond{(s\Gets A \PC{1/2} s\Gets B)}{(s=A)}{(s\Gets A)}
\]
operating on state space $\{A,B\}$. (In fact formula $\Psi$ expresses the
notorious $\textsf{AF}\,\textsf{AX}\,\AtB$
\cite{BvLTFS} in the temporal subset
\cite{aEBMfPTL} of \QTL, where $\VV.\AtB.s$ is defined to be $\Cond{1}{(s=B)}{0}$~.
Player 1 can force satisfaction of $\Psi$ with probability one in this game, since
the only path for which it fails (all $A$'s) occurs with probability zero; so de Alfaro' construction yields a \emph{different} formula $\phi$ such that $\Koz{\phi}{\VV} = 1$.

Yet $\Koz{\Psi}{\VV}$ for the original formula is only $1/2$, which is the value of the \emph{Stirling} game
played in this system. It is ``at each step, seek to maximise ($\Max$) the
payoff, depending on whether after the following step ($\NextTime{\KK}$) you
will accept $\AtB$ and terminate, or go around again ($X$).'' Note that the
decision ``whether to repeat after the next step''
is made \emph{before} that step is taken. (Deciding after the step
would be described by the formula $\CompNY{\mu}{X}{\NextTime{\KK}(\AtB \Max
X)}$.) The optimal strategy for \MAX\ is of course given by
$\Psi_{\overline{\textsf{\scriptsize atA}}} \Defs
\CompNY{\mu}{X}{\Cond{\NextTime{\KK}\AtB}{\textsf{atA}}{\NextTime{\KK}X}}$.

Finally, our result \Lem{memoriless} for memoriless strategies holds for all
$\QTL$ formulae, whereas (we believe) de Alfaro \EtAl\ treat only a subset,
those formulae encoding the automata used in their construction.

\bigskip
More recently, de Alfaro has given theorems for equivalence of game- and denotational interpretations of quantitative $\mu$-calculus formulae for ``discounted'' two-player games, provided the formulae are ``strongly deterministic'' \cite{QVaCvtMC}. \emph{Strongly deterministic} is a syntactic criterion that restricts to formulae that avoid the difference we illustrate above: that is, their game-value, as we define it, and their ``proportion of paths LTL-satisfying'' value (as above) are in agreement.%
\footnote{The restriction also excludes for example the case study of \Sec{LSM}, where our interest is genuinely in a game's \emph{minimax} value, rather than in the probability of satisfying an LTL specification. The special case treated in \Sec{s1135} can however be expressed in pCTL.}
\emph{Discounted} (turn-based) games, in our terms, are a special case of our Everett-style payoff states in which the probability of transition to \$\ is the complement $1-\alpha$ of the discount factor $\alpha$, as illustrated in \Fig{f1154}.

{\small
\bibliographystyle{plain}
\bibliography{probs,qMuGames}
}

\appendix
\section{Full proof of \Lem{l1227} from \Sec{s1812}}\label{s1227a}

\Lem{l1227} \emph{Logic/game equivalence of fixed-strategy interpretations} states that
\begin{quote}
For all closed \QTL\ formulae $\phi$, valuations $\VV$, states $s$ and strategies
$\Mins,\Maxs$, we have
\begin{equation}\label{e1403}
 \Exp{\Tree{\phi}{\VV}{\Mins,\Maxs}.s} \Val
 \quad=\quad
 \KozS{\phi}{\VV}{\Mins,\Maxs}.s ~,
\end{equation}
where $\Val$ is given by \Def{d0920}, the tree-building semantic function $\Tree{\cdot}{}{}$ is as given in \Fig{f0856}, and the strategy-extended denotational semantics $\KozS{\cdot}{}{}$ is given at \Fig{f0949} below.
\end{quote}

\begin{proof}\rm
We use structural induction over a stronger hypothesis including explicit paths (at \Eqn{e1402} below), straightforward except when least- or greatest fixed-points generate infinite trees; in each case the current formula will be $\phi$, and its constituent formula(e) will be $\Phi$ (with primes if necessary).  During the proof we formalise the use of strategies in both interpretations, extending both semantic functions with a ``path'' argument of type $\Pi$, say, which records the steps as the formula is decomposed and is used in the $\Max$($\Min$) case as the argument to the strategy $\Maxs$($\Mins$).

\paragraph{The tree construction} within the inductive argument introduces two new features: (a) that the current tree may in fact be a subtree, depending from some path $\pi\In \Pi$ in the overall tree corresponding to the original formula; and (b) that even though the whole tree is built from a closed formula, we must consider free variables in the inductive argument because it descends into the body of fixed-points.

The first feature (a) affects the use of the strategy functions: when resolving a $\Min$-choice, say, the path passed to the history-dependent minimising strategy $\Mins$ must be the path from the overall root, that is the current path within the subtree \emph{appended to} the path $\pi$ from which the whole subtree depends.%
%
Thus we supply a path as an extra argument to the tree-generating function, that is we write $\Tree{\phi}{\VV}{\Mins,\Maxs}.\pi.s$, following the convention that $\pi$ does not include the current position $(\phi,s)$.%
\footnote{The alternative approach of passing pre-determined strategy \emph{sequences} --- for example, an infinite sequence of Booleans each meaning ``go left'' or ``go right'' and consumed as it is used --- is not available to us.

Normally one argues that such sequences achieve full access to the history because, in pre-selecting say $\True$ or $\False$ for a given position in the strategy sequence, one has already made all the earlier selections --- and from those the formula/state that the current Boolean must deal with can in principle be determined. In our case the probabilistic choices are taken as the game is played, and the current formula/state cannot be predicted: thus the strategy \emph{functions} take an explicit path argument in order to look back and see how earlier probabilistic choices were resolved.}
If $\pi$ is omitted (as in the statement of the lemma) then it is taken to be the empty path $\langle\rangle$.

\noindent\makebox[0pt][r]{$+$\quad}\hspace{\parindent}%
The explicit path argument also provides a neat formalisation of the colour operations: we simply let the colours be subscripted variables, creating colour $X_i$ from bound variable $X$ where $i$ is the length of the path $\pi$ at the point the fixed-point formula binding $X$ is encountered. Then to look up colour $X_i$ at some later point $\pi'$ extending $\pi$, we simply take the $i^\mathit{th}$ element of $\pi'$ --- it will contain a fixed-point formula --- and we construct $\Phi[X \mapsto X_i]$ for the formula retrieved.

Strategies achieve colour-insensitivity by ignoring the subscript, treating position $(X_i,s)$ as just $(X,s)$.

For the second feature (b) we assume that all free variables $X$ in the current formula are defined in the valuation $\VV$, taken to functions of type $\Pi\rightarrow S \rightarrow [0,1]$; note that these functions deliver \emph{real values}, not subtrees. If we encounter $X$ when building the tree from current path $\pi$ and state $s$, we look up the value $X$ in $\VV$ to get a function $f$, and then insert the leaf node $(f.\pi^+.s)$ directly into the tree at that point, where $\pi^+$ is path $\pi$ routinely extended (as in \Fig{f0949} for the Kozen semantics) with the current game position, in this case $(X,s)$. The intention is that the stored function $f$ ``short-circuits'' the continued play from $(X,s)$ after path $\pi^+$: it simply supplies the value directly.

Note that our extended tree-building looks up free variables $X$ in the valuation $\VV$ ultimately to give a real number $x$ which is inserted as a leaf-node $(x)$, whereas colours $X_i$ refer to position $i$ in the path $\pi$ to give a formula $\Phi_{[X \mapsto X_i]}$ from which the tree-building then continues. A summary of the process was shown in \Fig{f0856}, and the game was given in \Fig{f0951}).

\paragraph{The extended Kozen semantics} $\KozS{\phi}{\VV}{\Mins, \Maxs}.\pi.s$ also accepts strategy sequences $\Mins,\Maxs$ and a path argument $\pi$, and in the definitions the path argument is routinely extended step-by-step so that it simulates the path that would be encountered in the corresponding tree; see \Fig{f0949}. Again, an omitted path defaults to empty. 

\begin{Figure}
\begin{enumerate}
\item\label{i1506a} $\KozS{X}{\VV}{\Mins,\Maxs} \Wide{\Defs} \VV.X$ ~.
\item\label{i1322} $\KozS{\AAA}{\VV}{\Mins,\Maxs}.\pi.s \Wide{\Defs} \VV.\AAA.s$ ~.
\item
    $\KozS{\NextTime{\KK} \Phi}{\VV}{\Mins,\Maxs}.\pi.s
    \Wide{\Defs}
    \VV.\KK.s.\$ ~+~ \Exp{\VV.\KK.s}\KozS{\Phi}{\VV}{\Mins,\Maxs}.\pi^+$ ~.
\item[] 
\item\label{i1219a}
    $\KozS{\Phi' \Min \Phi'' }{\VV}{\Mins,\Maxs}.\pi
    \Wide{\Defs}
    \begin{array}[t]{l@{\quad}l}
     \KozS{\Phi'}{\VV}{\Mins,\Maxs}.\pi^+
     & \textrm{if $\Mins.\pi^+$} \\
     \KozS{\Phi''}{\VV}{\Mins,\Maxs}.\pi^+
     & \textrm{otherwise.}
    \end{array}$
\item[]
\item
    $\KozS{\Bchoice{\GGG}{\Phi'}{\Phi''}}{\VV}{\Mins,\Maxs}.\pi.s
    \Wide{\Defs}
    \begin{array}[t]{l@{\quad}l}
     \KozS{\Phi'}{\VV}{\Mins,\Maxs}.\pi^+.s
     & \textrm{if $\VV.\GGG.s$} \\
     \KozS{\Phi''}{\VV}{\Mins,\Maxs}.\pi^+.s
     & \textrm{otherwise.}
    \end{array}$
\item[]
\item\label{i1323}
 $\KozS{\CompNY{\mu}{X}{\Phi}}{\VV}{\Mins,\Maxs}.\pi
 \Wide{\Defs}
 \CompNY{\Lfp}{x}{\KozS{\Phi}{\VV_{[X \mapsto x]}}{\Mins,\Maxs}}.\pi^+$
\item\label{i1324}
 $\KozS{\CompNY{\nu}{X}{\Phi}}{\VV}{\Mins,\Maxs}.\pi
 \Wide{\Defs}
 \CompNY{\Gfp}{x}{\KozS{\Phi}{\VV_{[X \mapsto x]}}{\Mins,\Maxs}}.\pi^+$
\item[]
\item (Colours are not used in $\KozS{\cdot}{}{}$ semantics.)
\end{enumerate}

The extra argument $\pi$ is a sequence of game positions, called a \emph{path};
in each case path $\pi^+$ is defined to be $\pi$ extended with the game position $(\phi,s)$, where $\phi$ is the  entire formula on the left-hand side.

\bigskip
Note that in Case \ref{i1506a} the value $\VV.X$ retrieved from the environment is applied to the current path and state; in Case \ref{i1322} however, only the state is used.

\bigskip
The strategy functions $\Maxs,\Mins$ are passed the current path when required (in Clause \ref{i1219a}, where we give only the $\Min/\Mins$ case).

\bigskip
The type of $x$ in the fixed-point clauses (\ref{i1323},\ref{i1324}) is \emph{path} to \emph{state} to $[0,1]$.

\caption{\label{f0949} Path/strategy-extended Kozen semantics; compare \Fig{f1210}.}
\end{Figure}

\paragraph{The inductive argument} thus treats the stronger hypothesis which includes the above features; it is that for all \QTL\ formulae $\phi$, valuations $\VV$, paths $\pi$, states $s$ and strategies $\Mins,\Maxs$, we have
\begin{equation}\label{e1402}
 \Exp{\Tree{\phi}{\VV}{\Mins,\Maxs}.\pi.s} \Val
 \quad=\quad
 \KozS{\phi}{\VV}{\Mins,\Maxs}.\pi.s ~,
\end{equation}
provided all free variables in $\phi$ are mapped by $\VV$ to functions of type $\Pi \rightarrow S \rightarrow [0,1]$ and that all colours in $\phi$ are mapped to formulae by $\pi$.
Our original goal \Eqn{e1403} is the case of \Eqn{e1402} in which $\VV$ defines only language constants and $\pi$ is empty.%
\footnote{Recall that neither colours nor free variables appear in the original formula, which is why the specialisation of \Eqn{e1402} to \Eqn{e1403} is appropriate.}

\bigskip
We now give a representative selection of the cases in the inductive argument.

\paragraph{\underline{Base case} $\phi$ is $X$}---\quad
From \Fig{f0856} we have that the game-subtree $\Tree{X}{\VV}{\Mins,\Maxs}.\pi.s$ is just the tip $(\VV.X.\pi.s)$, which value we note from the typing of $\VV$ given just before \Eqn{e1402} is indeed a real $r$, say, in $[0,1]$; from \Def{d0920} of $\Val$ we then have that the left-hand side of \Eqn{e1402} has value $r$.

From Case \ref{i1506a} of \Fig{f0949} we have that the right-hand side is
\[
 \KozS{X}{\VV}{\Mins,\Maxs}.\pi.s
 \Wide{=}
 (\VV.X).\pi.s ~,
\]
\Ie\ is $r$ also.

\paragraph{\underline{Base case} $\phi$ is $\AAA$}---\quad
Here the game-subtree $\Tree{\AAA}{\VV}{\Mins,\Maxs}.\pi.s$ is just the tip $(\VV.\AAA.s)$, which is correctly typed because $\VV$ takes constant-expectation symbols to functions in $S \rightarrow [0,1]$. The path is ignored.

From Case \ref{i1322} of \Fig{f0949} we have that the right-hand side is
\(
 \KozS{\AAA}{\VV}{\Mins,\Maxs}.\pi.s
 =
 \VV.\AAA.s ~,
\)
in which again the path is ignored.

\paragraph{\underline{Inductive case} $\phi$ is $\NextTime{\KK}\Phi$}---\quad
The game-subtree $\Tree{\NextTime{\KK}\Phi}{\VV}{\Mins,\Maxs}.\pi.s$ has $(\phi,s)$ at its root, and is extended by a finite number of branches, one to each possible next state $s'$ plus one to the special payoff state $\$$. Beneath branch $s'$, which has probability $\VV.\KK.s.s'$, is the subtree $\Tree{\Phi}{\VV}{\Mins,\Maxs}.\pi^+.s'$ where $\pi^+$ is $\pi$ extended with $(\phi,s)$ to record the node just passed through; and branch $\$$, which has probability $1-\sum_{s'\In S}\VV.\KK.s.s'$, is terminated by $(y)$ where the real value $y \in [0,1]$ is the payoff $\VV.\KK.s.\$/(1-\sum_{s'\In S}\VV.\KK.s.s')$ as at \Eqn{e0946}.

We now have
\begin{Reason}
\Step{}{
	\Exp{\Tree{\phi}{\VV}{\Mins,\Maxs}.\pi.s} \Val
}
\Space
\Step{$=$}{
	\Exp{\Tree{\NextTime{\KK}\Phi}{\VV}{\Mins,\Maxs}.\pi.s} \Val
}
\Space
\WideStepR{$=$}{for $\Val\,'$ derived from $\Val$\,: see $(\dag)$ below}{
	\VV.\KK.s.\$
	\Wide{+}
	(\sum_{s' \In S}~
		\VV.\KK.s.s'
		\Times
		\Exp{\Tree{\Phi}{\VV}{\Mins,\Maxs}.\pi^+.s'} \Val\,'\hspace{3em})
}
\Space
\WideStepR{$=$}{$\Val\,',\Val$ prefix-insensitive: see $(\ddag)$ below}{
	\VV.\KK.s.\$
	\Wide{+}
	(\sum_{s' \In S}~
		\VV.\KK.s.s'
		\Times
		\Exp{\Tree{\Phi}{\VV}{\Mins,\Maxs}.\pi^+.s'} \Val\hspace{3em})
}
\Space
\StepR{$=$}{structural induction}{
	\VV.\KK.s.\$ 
	\Wide{+}
	(\sum_{s' \In S}~
		\VV.\KK.s.s'
		\Times
		\KozS{\Phi}{\VV}{\Mins,\Maxs}.\pi^+.s')
}
\Space
\Step{$=$}{
	\VV.\KK.s.\$
	\Wide{+}
	\Exp{\VV.\KK.s} \KozS{\Phi}{\VV}{\Mins,\Maxs}.\pi^+.s'
}
\Space
\StepR{$=$}{\Fig{f0949}}{
	\KozS{\NextTime{\KK}\Phi}{\VV}{\Mins,\Maxs}.\pi.s
}
\Step{$=$}{
	\KozS{\phi}{\VV}{\Mins,\Maxs}.\pi.s ~.
}
\end{Reason}
as required for this case.

\bigskip\noindent\makebox[0pt][r]{$\dag$\quad}%
In the deferred justifications we are simply using the way in which expected values operate over tree-based distributions. We note that the expected value $E$ of $\Val$ over a (sub-)tree $T$ is the sum over its immediate children $T_i$ of the expected value $E_i$ assigned to $T_i$ times the probability $p_i$ labelling the branch $i$ that leads to it: that is, $E = \sum_i p_i \times E_i$. The expected values for the children are calculated just as for the parent, except that as we examine each child on its own, from \emph{its} root, we must use the function $\Val\,'$ defined $\Val\,'.\pi' \Defs \Val.(\textrm{``$(\phi,s)$ followed by $\pi'$''})$ instead of the original $\Val$, to take account of the fact that we have passed through node $(\phi,s)$ before reaching the child.

\bigskip\noindent\makebox[0pt][r]{$\ddag$\quad}%
We can however exploit the nature of our particular $\Val$, that it is not affected by adding finite prefixes to its argument; thus we can immediately replace $\Val\,'$ by $\Val$ again.

From that point on, the calculation of expected values $\Exp{}\hspace{-.5em}\Val$ behaves as above.

\paragraph{\underline{Inductive case} $\phi$ is $\Phi' \Min \Phi''$}---\quad
The game-tree again has $(\phi,s)$ at its root, but is extended with a single probability-one branch leading either to $\Tree{\Phi'}{\VV}{\Mins,\Maxs}.\pi^+.s$ or $\Tree{\Phi''}{\VV}{\Mins,\Maxs}.\pi^+.s$ depending on whether $\Mins.\pi^+$ is $\True$ (take $\Phi'$) or $\False$ (take $\Phi''$). Note that the state is not changed, and that the strategy function is applied to $\pi^+$ (not $\pi$), so that it has access to the current formula and state.

\paragraph{\underline{Inductive case} $\phi$ is $\CompNY{\mu}{X}{\Phi}$}---\quad
Here in Fig.~\ref{f0949}(\ref{i1002}) we appeal to $\Sup$-continuity%
\footnote{Because we have both least- and greatest fixed points, the justification of this assumption is not the usual ``continuity is preserved by the operation of taking fixed points'': for example $\sqcup$-continuity is not necessarily preserved by $\nu$.

In fact we have analytic continuity, which over $[0,1]$ implies $\sqcup,\sqcap$ continuity, from \Lem{Everett}: see the remark about its being maintained inductively, at ($+$) in the proof.}
to write the right-hand side as a limit
\[
 \CompNY{\Sup}{n}{f^n.\Const{0}}.\pi^+.s
 \quad
 \begin{array}[t]{@{}l@{\quad}l}
  \textrm{where} & f.x \Defs \KozS{\Phi}{\VV_{[X \mapsto x]}}{\Mins,\Maxs} \\
  \textrm{and} & \Const{0}.\pi'.s' \Defs 0 ~\textrm{for all $\pi'$ and $s'$,}
 \end{array}
\]
after which we will show by mathematical induction that for all $n$, states $s'$ and all extensions $\pi'$ of $\pi^+$ we have
\begin{equation}\label{e1217}
 f^n.\Const{0}.\pi'.s'
 \Wide{=}
 \Exp{\Tree{\Phi_{[X \mapsto X_i]}}{\VV}{\Mins,\Maxs}.\pi'.s'} \Val^{\,X_i}_n
\end{equation}
for suitably defined approximants $\Val^{\,X_i}_n$ of $\Val$, where $X_i$ is the colour chosen at position $(\phi,s)$ during the tree-building when the fixed-point formula was encountered.%

Our overall conclusion will follow by taking limits on both sides, appealing to \emph{bounded monotone convergence} \cite{PaI} to distribute through \SmallExp\ on the right.

Define $\Val^{\,X_i}_n.\pi'$ for any path $\pi'$ to be just $\Val.\pi'$ provided $\pi'$ contains fewer than $n$ occurrences of colour $X_i$; if however $\pi'$ contains at least $n$ occurrences of $X_i$, define $\Val^{\,X_i}_n.\pi'$ to be zero instead. We have $\CompNY{\Sup}{n}{\Val^{\,X_i}_n} = \Val$ because for all $\pi'$ with only finitely many $X_i$ we have $\Val^{\,X_i}_n.\pi' = \Val.\pi'$ for large-enough $n$; and for those $\pi'$ with infinitely-many $X_i$ we have zero in both cases.%
\footnote{Here is where we use the fact that $\Val$ is defined to yield zero if a $\mu$-colour occurs infinitely often.}

We now give the proof of \Eqn{e1217}, by induction over $n$: in Case 0, both sides are zero.

In Case $n+1$, we reason that for all $s'$ and extensions $\pi'$ of $\pi^+$ we have
\begin{Reason}
\Step{}{
 f^{n+1}.\Const{0}.\pi'.s'
}
\Step{$=$}{
 f.(f^n.\Const{0}).\pi'.s'
}
\Space
\StepR{$=$}{definition $f$}{
 \KozS{\Phi}{\VV_{[X \mapsto f^n.\Const{0}]}}{\Mins,\Maxs}.\pi'.s'
}
\Space
\StepR{$=$}{structural induction\footnotemark}{
 \EXP{\Tree{\Phi}{\VV_{[X \mapsto f^n.\Const{0}]}}{\Mins,\Maxs}.\pi'.s'} \Val
}
\Space
\StepR{$=$}{
 \begin{tabular}[t]{ll}
  inductive appeal to \Eqn{e1217} --- 
  for all extensions $\pi''$ of $\pi'$, \\
  and states $s''$, define \\\\
  $g.\pi''.s''
  \Wide{\Defs}
  \EXP{\Tree{\Phi_{[X \mapsto X_i]}}{\VV}{\Mins,\Maxs}.\pi''.s''} \Val^{\,X_i}_n$ \\\\
  so that $g = f^n.\Const{0}$
 \end{tabular}}
{
 \EXP{\Tree{\Phi}{\VV_{[X \mapsto g]}}{\Mins,\Maxs}.\pi'.s'} \Val
}
\Space\Space
\StepR{$=$}{
 \begin{tabular}[t]{ll}
  Because $\Phi$ contains no $X_i$, 
  and $\pi'$ extends $\pi$, \\
  tree $\Tree{\Phi}{\VV_{[X \mapsto g]}}{\Mins,\Maxs}.\pi'.s'$
  made from them \\
  will contain no $X_i$'s either; \\
  thus replacing $\Val$ by $\Val^{\,X_i}_{n+1}$ \\
  will make no difference; \\
  see ($\dag$) below.
 \end{tabular}}
{
 \EXP{\Tree{\Phi}{\VV_{[X \mapsto g]}}{\Mins,\Maxs}.\pi'.s'} \Val^{\,X_i}_{n+1}
}
\Space
\StepR{$=$}{see ($\ddag$) below}{
 \EXP{\Tree{\Phi_{[X \mapsto X_i]}}{\VV}{\Mins,\Maxs}.\pi'.s'} \Val^{\,X_i}_{n+1} ~,
}
\end{Reason}
\footnotetext{Here is where we use the extended hypothesis \Eqn{e1402}, rather than the original \Eqn{e1403}, because $\Phi$ may contain free variable $X$. Also, we rely here on ``for all $\VV$'' being part of the inductive hypothesis, since we are using $\VV[X \mapsto f^n.\Const{0}]$.}
thus establishing the inductive case.

\bigskip\noindent\makebox[0pt][r]{$\dag$\quad}%
For the first deferred justification, we note that $X_i$'s can come from only three places: (1) from $\Phi$ itself (but $\Phi$ contains no $X_i$, since $X_i$ was fresh); (2) from the interior of formulae retrieved from $\pi$ by looking up \emph{other} colours in $\Phi$ (but $\pi$ contains no ``embedded'' $X_i$ either, again because it was fresh); or (3) from the subsequent creation of colours (but they themselves will be fresh, different from $X_i$, by construction --- guaranteed by the fact that the length of $\pi'$ exceeds the length of $\pi$ and that the length determines the subscript of the any newly-created colour).

That is the import of ``choose a fresh colour'' in the tree-building algorithm.

\bigskip\noindent\makebox[0pt][r]{$\ddag$\quad}%
For the final step we are claiming, roughly speaking, that at all the points in the constructed tree where $X$ occurs (left-hand side) or ``used to be'' (right-hand side, now replaced by $X_i$), the function $g$ was defined precisely so that it makes no difference to the integral
$\int\!\Val^{\,X_i}_{n+1}$ whether we
\begin{enumerate}
\item\label{i1024} look up variable $X$ in $\VV_{[X\mapsto g]}$ to get $g$, which applied to the path $\pi''$ and state $s''$ at that point gives a tip $(g.\pi''.s'')$ directly, or
\item[]
\item\label{i1025} look up colour $X_i$ in path $\pi''$, to recover the formula $\Phi_{[X\mapsto X_i]}$ and carry on building the tree below.
\end{enumerate}
That is, the value in the tip constructed at (\ref{i1024}) is exactly the value realised from the tree constructed at (\ref{i1025}) by the integral $\int\!\Val^{\,X_i}_{n+1}$.

\bigskip
In more detail: we are in fact relying on an elementary property of $\SmallExp F$ over game-trees, for general $F$. Take any game-tree $T$, and describe subtrees of it as pairs $\langle \pi, U \rangle$, where $U$ is (also) a game-tree and $\pi$ is the path leading from the root of $T$ to just before the root of $U$. Let $T[\langle \pi, U \rangle \mapsto V]$ be the tree resulting from replacing that entire subtree by another tree $V$. We then have that
\begin{equation}\label{e1056}
 \Exp{T[\langle \pi, U \rangle \mapsto V]} F
 \hspace{4em}
 \Wide{=}
 \Exp{T}F
 ~-~
 \Exp{U} F_\pi
 ~+~
 \Exp{V} F_\pi ~,
\end{equation}
where $F_\pi.\pi' ~\Defs~  F.(\pi \Concat \pi')$ for all $\pi'$.%
\footnote{We write $\Concat$ for path concatenation.}
In effect, on the right we subtract the contribution made by $U$ and then add back the contribution made by $V$, but in each case we use $F_\pi$ over the sub-tree to compensate for the fact that its contribution is made \emph{within} (\IE\ at $\pi$) the overall tree $T$.
Furthermore, the above holds for any countable pairwise-disjoint set of such substitutions done simultaneously.%
\footnote{We require that the set of all paths affected is measurable, which is why we require countability of the subtrees.}

\bigskip
Now for the final step in our proof above we reason backwards, from the last expression --- call it  $[\ddag]$ --- to the second-last, $[\dag]$, using an instantiation of \Eqn{e1056}.
We unify $[\ddag]$ and the first term on the right-hand side of \Eqn{e1056} by choosing function $F$ to be $\Val^{\,X_i}_{n+1}$, and the tree $T$ to be $\Tree{\Phi_{[X \mapsto X_i]}}{\VV}{\Mins,\Maxs}.\pi'.s'$.

Now Tree $T$ contains (at most) a countable number, $k$-indexed say, of ``first encounter of $X_i$ from the root of $T$\,'' positions $(X_i, s^k)$, and each is the final element of some path $\pi^k$ containing no other $X_i$; below each $\pi^k$ is some subtree $U^k$, which from our tree-construction procedure we know will be
\begin{equation}\label{e1120}
 \Tree{\Phi_{[X \mapsto X_i]}}{\VV}{\Mins,\Maxs}.(\pi' \Concat \pi^k).s^k ~,
\end{equation}
since $\Phi_{[X \mapsto X_i]}$ is what is returned when we look up colour $X_i$ and $\pi' \Concat \pi^k$ is the overall path that leads to this point.
(Refer Case \ref{i1107} of \Fig{f0856}.)
One-by-one we will use these $U^k$'s as $U$ in countably-many applications of \Eqn{e1056}. 

For each $k$ the function $F_\pi$ in \Eqn{e1056}, which we will call $F_{\pi^k}$, will be $(\Val^{\,X_i}_{n+1})_{\pi^k}$ because of our choice above of $F$. But that is just $\Val^{\,X_i}_n$, because $\pi^k$ contains exactly one $X_i$ (at its end) which ``uses up'' the $+1$ in the subscript $n{+}1$. Thus for each $k$ the
second term on the right of \Eqn{e1056} is the integral of $\Val^{\,X_i}_n$ taken over Tree \Eqn{e1120}, \emph{viz.}
\begin{equation}\label{e1110}
 x^k
 \Wide{\Defs}
 \EXP{\Tree{\Phi_{[X \mapsto X_i]}}{\VV}{\Mins,\Maxs}.(\pi' \Concat \pi^k).s^k} \Val^{\,X_i}_n ~.
\end{equation}

\bigskip
Now for the the third term we choose the $V^k$ (to replace $U^k$)
to be the trivial subtree comprising just a tip $(x^k)$; that makes $\Exp{(x^k)} \Val^{\,X_i}_n$ just $x^k$ again.

\bigskip
With the second and third terms in \Eqn{e1056}  equal, the first term on its own (which we recall is $[\ddag]$) equals the left-hand side.
Figures \ref{f1901} and \ref{f1900} illustrate the trees occurring in the left- and right-hand sides of \Eqn{e1056}.
\begin{Figure}
\begin{center}
\epsfxsize=20em \epsfbox{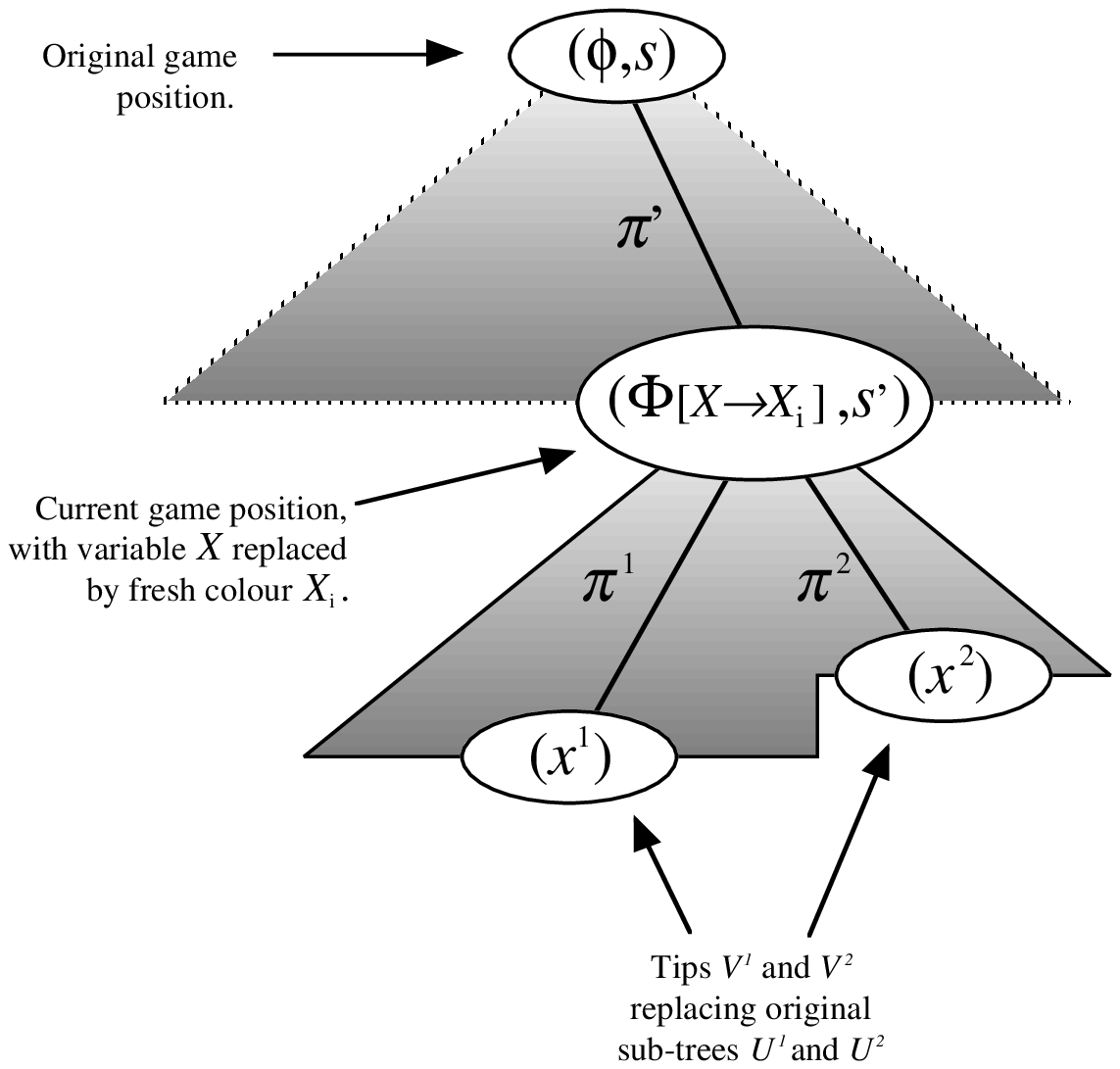}
\end{center}
\caption{Left-hand side of \Eqn{e1056}: tree $T$ after subtrees $U^k$ replaced by tips $V^k$.}
\label{f1901}
\end{Figure}
\begin{Figure}
\begin{center}
\epsfxsize=30em \epsfbox{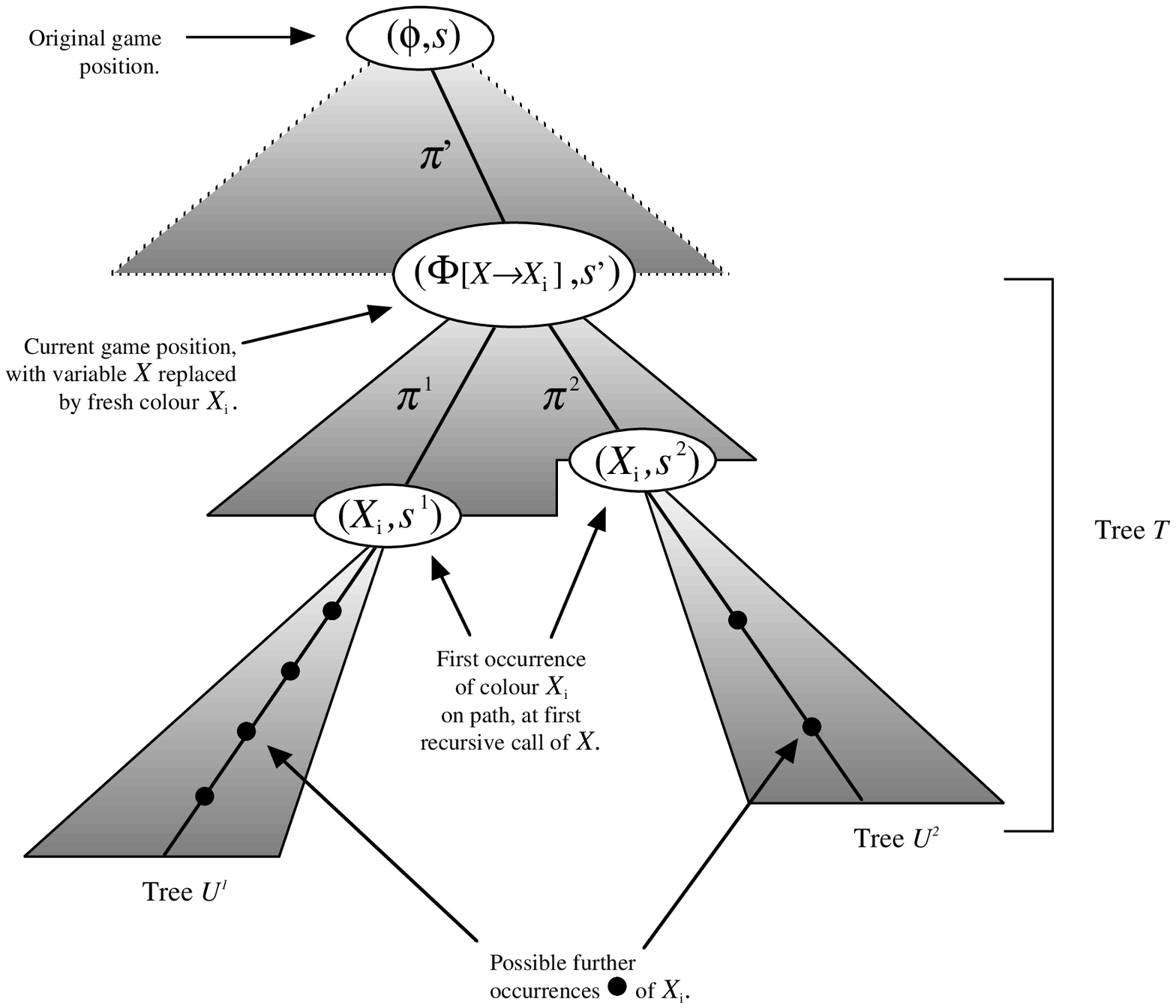}
\end{center}
\caption{Right-hand side of \Eqn{e1056}: tree $T$ before subtrees $U^k$ replaced by tips $V^k$.}
\label{f1900}
\end{Figure}

\bigskip
We will now show that the left-hand side of \Eqn{e1056} is equal to $[\dag]$.
The tree used there (\Fig{f1901}) is
\[
 T[\langle \pi^0, U^0 \rangle \mapsto V^0,\langle \pi^1, U^1 \rangle \mapsto V^1 \cdots]~,
\]
--- \IE\ the result of all the $k$-indexed substutitions done simultaneously --- and each $V^k$ is just the tip $(x^k)$. But the tree  $\Tree{\Phi}{\VV_{[X \mapsto g]}}{\Mins,\Maxs}.\pi'.s'$ used in $[\dag]$ is the same except that it contains the tip $(g.\pi^k.s^k)$ at those places. (The places agree because they are both determined by the occurrences of $X$ in the original formula $\Phi$.)

Comparison of the definition of $g$ (at $[\dag]$) --- noting its arguments at each $k$ will be $\pi'' \Defs \pi' \Concat \pi^k$ and $s'' \Defs s^k$ --- and the definition of $x^k$ (at \Eqn{e1110}) shows those tip-values to be equal.

\bigskip
That concludes our justification of the final step above, and of our inductive proof of \Eqn{e1217} as a whole.

\bigskip
Using \Eqn{e1217} we finish off the proof of this case as follows.
Choose path $\pi^+$ itself and state $s$; then with \emph{bounded monotone convergence} we have
\begin{Reason}
\Step{}{
 \CompNY{\Sup}{n}{f^n.\Const{0}}.\pi^+.s
}
\Space
\StepR{$=$}{from \Eqn{e1217} in the special case $\pi' = \pi^+$\,\footnotemark}{
 \CompNY{\Sup}{n}{\Exp{\Tree{\Phi_{[X \mapsto X_i]}}{\VV}{\Mins,\Maxs}.\pi^+.s} \Val^{\,X_i}_n}
}
\Space
\StepR{$=$}{bounded monotone convergence $(\ast)$}{
 \EXP{\Tree{\Phi_{[X \mapsto X_i]}}{\VV}{\Mins,\Maxs}.\pi^+.s} \CompNY{\Sup}{n}{\Val^{\,X_i}_n}
}
\Space
\StepR{$=$}{$\CompNY{\Sup}{n}{\Val^{\,X_i}_n} = \Val$}{
 \Exp{\Tree{\Phi_{[X \mapsto X_i]}}{\VV}{\Mins,\Maxs}.\pi^+.s} \Val
}
\Space
\StepR{$=$}{
 \begin{tabular}{ll}
  tree-building step for $\mu$ (backwards); \\
  $X_i$ looks up $\Phi_{[X \mapsto X_i]}$ in $\pi^+$
 \end{tabular}
}{
 \Exp{\Tree{\phi}{\VV}{\Mins,\Maxs}.\pi.s} \Val ~.
}
\end{Reason}
where the final step is the one in which colour $X_i$ was generated.
\footnotetext{The equality \Eqn{e1217} is for all extensions $\pi'$ of $\pi^+$ because of its inductive proof: the stronger hypothesis is used when defining $g$.}
 
\paragraph{\underline{Inductive case} $\phi$ is $\CompNY{\nu}{X}{\Phi}$}---\quad
This case is essentially the same as the $\mu$-case --- we define the truncated valuations $\Val^{\,X_i}_n.\pi'$ as before except that paths $\pi'$ with at least $n$ occurrences of $X_i$ are taken to one (rather than to zero).

A small complication however occurs in the use of bounded monotone convergence, which requires the sequence of valuations to be monotone non-decreasing: at the point corresponding to $(\ast)$ above we would in this case be arguing that
\[
 \CompNY{\Inf}{n}{\Exp{\Tree{\Phi_{[X \mapsto X_i]}}{\VV}{\Mins,\Maxs}.\pi^+.s} \Val^{\,X_i}_n}
 \Wide{\Wide{\Wide{\quad=}}}
 \EXP{\Tree{\Phi_{[X \mapsto X_i]}}{\VV}{\Mins,\Maxs}.\pi^+.s} \CompNY{\Inf}{n}{\Val^{\,X_i}_n}~,
\]
where the terms are non-\emph{increasing}. Since all the terms lie in $[0,1]$ however, we can deal with it by subtracting from one throughout, before and after.

\end{proof}

\section[Memoriless strategies suffice]{\begin{tabular}[t]{@{}l}
Memoriless strategies suffice \\
over a finite state space
\end{tabular}} \label{s1650}

We show that for any formula $\phi$, possibly including $\Min$ and $\Max$
strategy operators, there are specific state predicates (collected into tuples
$\UnderB$ and $\OverB$) that can replace the strategy operators without
affecting the value of the formula.
The inductive proof is straightforward except for replacement of $\Max$ within $\mu$
(and, dually, replacement of $\Min$ within $\nu$). For this $\OverB$/$\mu$ case we need
several technical lemmas and definitions; the other cases are set out at \cite{proofs}.

Because the argument in this section is mainly over properties of real-valued functions, we shift to a more mathematical style of presentation. Variables $f,g,\ldots$ denote (Curried) functions of type expectation(s) to expectation, and $w,x,\ldots$ are expectations in $\E{S}$. For function $f$ of one argument we write $\mu.f$ for its least fixed-point.

\begin{definition}\label{d1156} {\rm Almost-linear} --- \quad Say that an expectation-valued
function $f$ of possibly several expectation arguments $x,y,\cdots,z$ is
\emph{almost-linear} if it can be written in the form
\begin{equation}\label{e1206}
 f.x.y\cdots.z \Wide{\Defs} w + g.x + h.y + \cdots + i.z ~,
\end{equation}
where $w$ is an expectation and $g, h, \cdots, i$ are \emph{linear} expectation-valued functions of their single
arguments.
\end{definition}

\begin{lemma}\label{l1204} Every $\Min$/$\Max$-free formula $\phi$, possibly containing
free expectation variables $X,Y, \ldots, Z$, denotes an almost-linear function of
the values assigned to those arguments.

\begin{proof} (sketch)
What we are claiming is that the function
\[
 f.x.y\cdots.z \Wide{\Defs} \Koz{\phi}{\VV[x,y\cdots z/X,Y\cdots Z]}
\]
can be written in the form given on the right at \Eqn{e1206}, provided $\phi$
contains no $\Min$ or $\Max$. This is a straightforward structural induction
over $\phi$, given in full at \cite{proofs}.
\end{proof}
\end{lemma}

\begin{definition}\label{d1301} {\rm Almost less-than} --- \quad For non-negative reals $a,b$, write $a \Lle b$ for $a > 0 \Implies a < b$; write the same for the pointwise-extended relation
over expectations. Note that $a < b$ implies $a \Lle b$ implies $a \leq b$ on this domain.
\end{definition}

\begin{definition}\label{d1216} {\rm $\OK$ functions} --- \quad Say that an expectation-to-expectation function $f$ of one argument is
$\OK$ if for all expectations $x$ with $x \Lle f.x$ we have that $ x \leq \mu.f$ ~.
\end{definition}

\begin{lemma}\label{l1228} If $f$ is almost-linear then $f$ is $\OK$ in each argument separately.

\begin{proof} See Appendix \ref{a1250}.
\end{proof}
\end{lemma}

\begin{lemma}\label{l1229} All $\Min$/$\Max$-free formulae $\phi$ denote $\OK$
functions of their free expectations $X,Y, \cdots ,Z$ taken separately.

\begin{proof}
 Lemmas \ref{l1204} and \ref{l1228}.
\end{proof}
\end{lemma}

The following result forms the core of Everett's argument
\cite{RG}; note it does not depend on $f$'s being $\OK$.

\begin{lemma}\label{l1234} For any monotonic and continuous%
\footnote{\label{f1106}This is continuity in the usual sense in analysis; see Footnote \ref{f1107}. With monotonicity we have $\Max$-continuity for $f$ as well.}
function $f$ over expectations, and any $\varepsilon > 0$, there is an expectation $x$ such that
\begin{eqnarray}
 x &\Lle& f.x \label{e1040b} \\
 \textrm{and\hspace{5em}}
  \Lfp.f - \Const{\varepsilon} &\leq& x ~,\label{e1040a}
\end{eqnarray}
where $\Const{\varepsilon}$ is the everywhere-$\varepsilon$ expectation. That is, we can find an almost-increased-by-$f$ expectation $x$ that approaches $\Lfp.f$ as closely as we please from below.

\begin{proof} Define a subset $T$ of the state space $S$ by
\begin{equation}\label{e1037}
 T \Wide{\Defs} \SetCompYN{s\In S}{f.\Const{0}.s = \Lfp.f.s} ~,
\end{equation}
so that the subset $T$ is ``the termination set for $f$,'' comprising those states at which $f$ reaches its fixed-point in just one step. Because $S$ is finite we can proceed by induction decreasing over these sets $T$ determined by $f$, with the base case therefore being when $T$ is all of $S$.

We strengthen%
\footnote{\label{n1122}The extra condition $x \leq \Lfp.f$ is used at Footnote \ref{n1123} in the subsidiary \Lem{l1758} below.}
Condition \Eqn{e1040a} of the inductive hypothesis to read
\par\bigskip
\hfill $\Lfp.f - \Const{\varepsilon} \Wide{\leq} x \Wide{\leq} \Lfp.f$~. \hfill {\rm (\ref{e1040a}a)}

\paragraph{\underline{Case} $T=S$}---\quad
Define $x \Defs (\Lfp.f - \Const{\varepsilon}) \Max \Const{0}$ so that {\rm (\ref{e1040a}a)} is satisfied trivially. Since $\varepsilon>0$ we have also that $x \Lle \Lfp.f$, and then from $T=S$ and monotonicity of $f$ we reason
\[
 x \Wide{\Lle} \Lfp.f \Wide{=} f.\Const{0} \Wide{\leq} f.x ~,
\]
sufficient for \Eqn{e1040b}.

\paragraph{\underline{Case} $T \subset S$}---\quad
Pick $s^\ast$ from $S-T$, and for all $x$ define
\[
 f^\ast_v.x \Wide{\Defs} f.x_{[s^\ast \mapsto v]} ~,
\]
that is the expectation that agrees with $f.x$ everywhere except possibly at $s^\ast$ where it takes the value $v$ instead.%
\footnote{In the following argument we hold $s^\ast$ fixed, which is why to avoid clutter we can omit it from the notation $f^\ast_v$. We will however vary $v$.} 
Define also $v^\ast \Defs \Lfp.f.s^\ast$, and note that $v^\ast > 0$ because otherwise we would have $s^\ast \in T$.%
\footnote{\label{n1003}This fact is used at Footnote \ref{n1004}.}

We begin by showing two things about $f^\ast_v$. The first (a) is that ``the termination set for $f^\ast_v$'' --- that is $T^\ast_v \Defs \SetCompYN{s\In S}{f^\ast_v.\Const{0}.s = \Lfp.f^\ast_v.s}$ --- is a strict superset of $T$ when $v \leq v^\ast$, which will allow an appeal to the inductive hypothesis. The second (b) is that the function $\Lfp.f^\ast_v$ of $v$ approaches $\Lfp.f$ as $v$ approaches $v^\ast$ from below, and attains it in the limit. 

\bigskip
To show (a) we assume $v \leq v^\ast$ and note first that
\begin{equation}\label{e1146}
 \Lfp.f^\ast_v \Wide{\leq} \Lfp.f ~,
\end{equation}
because
\[
 \begin{array}{r@{\quad}ccccc@{\quad}l}
  & f^\ast_v.(\Lfp.f).s &\Wide{=}& f.(\Lfp.f).s &\Wide{=}& \Lfp.f.s
   & \textrm{for $s \neq s^\ast$} \\
  \textrm{and}
   & f^\ast_v.(\Lfp.f).s^\ast &\Wide{=}& v &\Wide{\leq}& \Lfp.f.s^\ast & \textrm{by assumption.}
 \end{array}
\]
That establishes $f^\ast_v.(\Lfp.f) \leq \Lfp.f$, sufficient for \Eqn{e1146} by the least-fixed-point property.%
\footnote{The \emph{least-fixed-point property} states that $f.x \leq x$ implies $\Lfp.f \leq x$ for any monotonic $f$ over a \emph{cpo}.}

From \Eqn{e1146} we show $T \cup \{s^\ast\} \subseteq T^\ast_v$ by considering two cases:
\begin{description}
\item[Case $s \in T$:] We have
\begin{Reason}
\Step{}{
 f^\ast_v.\Const{0}.s
}
\StepR{$\leq$}{monotonicity}{
 f^\ast_v.(\Lfp.f^\ast_v).s
}
\StepR{$=$}{fixed point}{
 \Lfp.f^\ast_v.s
}
\StepR{$\leq$}{\Eqn{e1146}}{
 \Lfp.f.s
}
\StepR{$=$}{$s \in T$}{
 f.\Const{0}.s
}
\StepR{$=$}{$s \neq s^\ast$}{
 f^\ast_v.\Const{0}.s~.
}
\end{Reason}
\item[Case $s =s^\ast$:] We have
\begin{Reason}
\Step{}{
 f^\ast_v.\Const{0}.s^\ast
}
\StepR{$=$}{definition $f^\ast_v$}{
 v
}
\StepR{$=$}{again definition $f^\ast_v$}{
 f^\ast_v.(\Lfp.f^\ast_v).s^\ast
}
\StepR{$=$}{fixed point}{
 \Lfp.f^\ast_v.s^\ast~.
}
\end{Reason}
\end{description}
Thus when $v \leq v^\ast$ we have $T \subset T \cup \{s^\ast\} \subseteq T^\ast_v$, which establishes (a).

\bigskip
To show (b) we note first that $\Lfp.f^\ast_v$ is a continuous function of $v$ as $v$ increases.%
\footnote{This general result --- continuity of fixed-points --- requires in this case that $f^\ast_v$ is $\Max$-continuous in $v$ and that each $f^\ast_v$ is itself $\Max$-continuous, the former trivial and the latter following from $\Max$-continuity of $f$. It gives continuity over directed sets of $v$, which we have because $v$ is increasing.}
Thus we need only demonstrate that $\Lfp.f^\ast_{v^\ast} = \Lfp.f$, since \Eqn{e1146} already shows that $\Lfp.f^\ast_v$ is below $\Lfp.f$ for $v \leq v^\ast$. Again we consider two cases (and appeal to \Eqn{e1146} itself in the second case):
\begin{description}
\item[Case $s \neq s^\ast$:] We have
$f.(\Lfp.f^\ast_{v^\ast}).s ~=~ f^\ast_{v^\ast}.(\Lfp.f^\ast_{v^\ast}).s ~=~ \Lfp.f^\ast_{v^\ast}.s$~.
\item[Case $s =s^\ast$:] We have
\begin{Reason}
\Step{}{
 f.(\Lfp.f^\ast_{v^\ast}).s^\ast
}
\StepR{$\leq$}{by \Eqn{e1146} in the special case $v=v^\ast$}{
 f.(\Lfp.f).s^\ast
}
\StepR{$=$}{fixed point}{
 \Lfp.f.s^\ast
}
\StepR{$=$}{definition $v^\ast$}{
 v^\ast
}
\StepR{$=$}{definition $f^\ast_{v^\ast}$}{
 f^\ast_{v^\ast}.(\Lfp.f^\ast_{v^\ast}).s^\ast
}
\StepR{$=$}{fixed point}{
 \Lfp.f^\ast_{v^\ast}.s^\ast ~.
}
\end{Reason}
\end{description}
Thus $f.(\Lfp.f^\ast_{v^\ast}) \leq \Lfp.f^\ast_{v^\ast}$, whence by the least-fixed-point property we have $\Lfp.f \leq \Lfp.(f^\ast_{v^\ast})$ which --- with \Eqn{e1146} again in the case $v = v^\ast$ --- gives the equality we need and establishes (b).

\bigskip
With (a) and (b) secure, we proceed to the main proof: we make a particular choice of $v$ and appeal to the induction hypothesis in respect of $f^\ast_v$ to find an expectation close to $\Lfp.f^\ast_v$, and we then show how to derive from that a suitable expectation $x$ satisfying (\ref{e1040b},\ref{e1040a}a) as required for the function $f$ in this case.

We choose $v$ first. From (b) we can choose $v< v^\ast$ to achieve%
\footnote{\label{n1004}Recall Footnote \ref{n1003} to see this is possible. In fact only $v \leq v^\ast$ is needed here, in the main proof; the strictness of the inequality is used at Footnote \ref{n1018} in \Lem{l1758} below.}
\begin{equation}\label{e1410}
 \Lfp.f - \Const{\varepsilon_1} \Wide{\leq} \Lfp.f^\ast_v \Wide{\leq} \Lfp.f
\end{equation}
for any $\varepsilon_1 < \varepsilon$ we please.%
\footnote{Here we use finiteness of the state space, since the one $\varepsilon_1$ applies for all states.}

Now we appeal to the induction hypothesis: since $T \subset T^\ast_v$, for any $0 < \varepsilon_2 \leq \varepsilon - \varepsilon_1$ we can find an $x^{\varepsilon_2}_v$ satisfying
\begin{eqnarray}
 \Lfp.f^\ast_v - \Const{\varepsilon_2} &\Wide{\leq}& x^{\varepsilon_2}_v \Wide{\leq} \Lfp.f^\ast_v \label{e1225aa} \\
 \textrm{and\hspace{5em}}
  x^{\varepsilon_2}_v &\Wide{\Lle}& f^\ast_v.x^{\varepsilon_2}_v ~. \label{e1225bb}
\end{eqnarray}
From that we have immediately
\begin{Reason}
\Step{}{
 \Lfp.f - \Const{\varepsilon}
}
\StepR{$\leq$}{choice of $\varepsilon_2$}{
 \Lfp.f - \Const{\varepsilon_1} - \Const{\varepsilon_2}
}
\StepR{$\leq$}{left-hand inequality at \Eqn{e1410}}{
 \Lfp.f^\ast_v - \varepsilon_2
}
\StepR{$\leq$}{left-hand inequality at \Eqn{e1225aa}}{
 x^{\varepsilon_2}_v
}
\StepR{$\leq$}{right-hand inequality at \Eqn{e1225aa}}{
 \Lfp.f^\ast_v
}
\StepR{$\leq$}{left-hand inequality at \Eqn{e1410}}{
 \Lfp.f ~,
}
\end{Reason}
which is our (\ref{e1040a}a) if we take $x$ to be $x^{\varepsilon_2}_v$.

All that remains is \Eqn{e1040b}, for which we require $x^{\varepsilon_2}_v \Lle f.x^{\varepsilon_2}_v$ --- and indeed that holds trivially everywhere except possibly at $s^\ast$: for if $s \neq s^\ast$ we have from \Eqn{e1225bb} that
\begin{equation}\label{e1803}
 x^{\varepsilon_2}_v.s
 \Wide{\Lle}
 f^\ast_v.x^{\varepsilon_2}_v.s
 \Wide{=}
 f.x^{\varepsilon_2}_v.s
 ~.
\end{equation}

Thus all we are left with is to show that $x^{\varepsilon_2}_v.s^\ast ~\Lle~ f.x^{\varepsilon_2}_v.s^\ast$, which unfortunately is not true for all $x^{\varepsilon_2}_v$ satisfying (\ref{e1225aa},\ref{e1225bb}). But, as we demonstrate in the technical \Lem{l1758} proved in \App{s1734} below, for any $\varepsilon_2>0$ it is possible to find an $\varepsilon^\ast_2$ with $0<\varepsilon^\ast_2 \leq \varepsilon_2$ which retains the properties (\ref{e1225aa},\ref{e1225bb},\ref{e1803}) above and satisfies $x^{\varepsilon^\ast_2}_v.s^\ast ~\Lle~ f.x^{\varepsilon^\ast_2}_v.s^\ast$ as well --- and which thus completes the proof.
\end{proof}
\end{lemma}

We can now sketch the proof of the main result of this section.
\begin{lemma}\label{Everett} {\rm Fixed strategies suffice} --- \quad
For any formula $\phi$, possibly containing strategy operators
$\Min$/$\Max$, and valuation $\VV$, there are state-predicate
tuples $\UnderB$/$\OverB$ --- possibly depending on $\VV$ --- such that
\[
  \Koz{\phi_{\UnderB}}{\VV}
  \Wide{=}
  \Koz{\phi}{\VV}
  \Wide{=}
  \Koz{\phi_{\OverB}}{\VV} ~.
\]

\begin{proof} (sketch)
We give only the $\mu$-case of an otherwise
straightforward induction over the size of $\phi$; a full proof may be found at
\cite{proofs}.

Suppose we are considering the case where $\phi$ is a least-fixed-point $\CompNY{\mu}{X}{\Phi}$. Let $f$ be the function
denoted by $\Phi$ with respect to a single expectation-valued
argument $x$ supplied for the variable $X$, with the values of
any other free variables in $\Phi$ fixed by the environment $\VV$;
for any $\UnderB,\OverB$ let functions $f_{\OverB}$ and $f_{\UnderB,\OverB}$
be derived similarly from $\Phi_{\OverB}$ and $\Phi_{\UnderB,\OverB}$.

\paragraph{\rm Case $\UnderB$}---\quad We must show $\mu.f = \mu.f_{\UnderB}$ for some $\UnderB$;%
\footnote{Note that $\CompNY{\mu}{X}{\Phi}_{\UnderB}$ is the same as
$\CompNY{\mu}{X}{\Phi_{\UnderB}}$ --- it is syntactic substitution ---
so that $\mu.(f_{\UnderB})$ is indeed the correct denotation.}
note that $\mu.f \leq \mu.f_{\UnderB}$ trivially, since $f \leq f_{\UnderB}$.
Since $\Phi$ is smaller in size than $\phi$, our inductive hypothesis provides for any $x$ a $\UnderB_x$ so that
$f_{\UnderB_x}.x = f.x$; take $x = \mu.f$ and therefore choose
$\UnderB$ so that $f_{\UnderB}.(\mu.f)  = f.(\mu.f) = \mu.f$. Thus $\mu.f$ is a fixed-point
of $f_{\UnderB}$, whence immediately $\mu.f_{\UnderB} \leq \mu.f$.

\paragraph{\rm Case $\OverB$}---\quad In this case must show $\mu.f = \mu.f_{\OverB}$ for some $\OverB$; again it is trivial that $\mu.f_{\OverB} \leq \mu.f$ for any $\OverB$.

For the other direction,
in fact we show that for any $\varepsilon > 0$ there is a $\OverB_\varepsilon$ such that
$\mu.f_{\OverB_\varepsilon} \geq \mu.f - \varepsilon$ --- whence the existence of a single
$\OverB$ satisfying $\mu.f_{\OverB} \geq \mu.f$ follows from the finiteness of
the state space (since the set of possible strategy tuples for this $f$ is therefore finite as
well, and so there must be one that works for all $\varepsilon$).

\noindent\makebox[0pt][r]{$+$\quad}\hspace{\parindent}%
Because we know inductively that $f$ is a \emph{minimax}%
\footnote{An argument similar to that used in \Lem{l1322} makes this explicit.}
over strategy tuples $\UnderB',\OverB'$ of almost-linear functions $f_{\UnderB',\OverB'}$,
that those functions are continuous by construction, and that the \emph{minimax} is finite because
there are only finitely many strategy tuples $\UnderB',\OverB'$ for this $f$,
we know that $f$ is continuous itself, and by \Lem{l1234} we therefore have an expectation
$x_\varepsilon$ with
\begin{equation}\label{e1239a}
 \mu.f - \varepsilon \leq x_\varepsilon 
 \quad \quad \textrm{and} \quad \quad
 x_\varepsilon \Lle f.x_\varepsilon ~. 
\end{equation}
To get our result we need only show in addition that $x_\varepsilon \leq \mu.f_{\OverB_\varepsilon}$ for some $\OverB_\varepsilon$.

From our inductive hypothesis we can choose $\OverB_\varepsilon$ so that $f.x_\varepsilon = f_{\OverB_\varepsilon}.x_\varepsilon$, whence from \Eqn{e1239a} we have $x_\varepsilon \Lle f_{\OverB_\varepsilon}.x_\varepsilon$. 
But in fact $f_{\OverB_\varepsilon}$ is $\OK$ (see below), so
from \Def{d1216} we have
$x_\varepsilon \leq \mu.f_{\OverB_\varepsilon}$ and we are done.

\begin{center}\rule{5cm}{.01cm}\end{center}

To see that $f_{\OverB_\varepsilon}$ is $\OK$, we apply the argument of Case $\UnderB$,%
\footnote{That argument makes an appeal to the inductive hypothesis in respect of $\Phi_{\OverB_\varepsilon}$, a smaller formula than $\phi$. Note however it is not a subformula of $\phi$, which is why we do not use structural induction.}
which gives us a $\UnderB'$ with $\mu.f_{\OverB_\varepsilon} = \mu.f_{\UnderB',\OverB_\varepsilon}$. Now consider any $x$ such that $x \Lle f_{\OverB_\varepsilon}.x$~.

Since $f_{\OverB_\varepsilon}.x \leq f_{\UnderB',\OverB_\varepsilon}.x$ we have $x \Lle f_{\UnderB',\OverB_\varepsilon}.x$ also --- but we recall from
\Lem{l1229} that     
$f_{\UnderB',\OverB_\varepsilon}$ is $\OK$. Hence
$x \leq \mu.f_{\UnderB',\OverB_\varepsilon} = \mu.f_{\OverB_\varepsilon}$~,
and $f_{\OverB_\varepsilon}$ is $\OK$ as well.
\end{proof}
\end{lemma}

\section{A technical lemma used in \App{s1650}} \label{s1734}
\begin{lemma}\label{l1758} In this technical lemma we continue the notation established within the proof of \Lem{l1234}; we show that for any $\varepsilon_2>0$ there is an $\varepsilon^\ast_2$ with  $0 < \varepsilon^\ast_2 \leq \varepsilon_2$ and an $x^{\varepsilon^\ast_2}_v$ that together retain the properties (\ref{e1225aa},\ref{e1225bb},\ref{e1803}) and in addition satisfy $x^{\varepsilon^\ast_2}_v.s^\ast ~\Lle~ f.x^{\varepsilon^\ast_2}_v.s^\ast$, as we require above.
\begin{proof}
In fact we find $\varepsilon^\ast_2$ and $x^{\varepsilon^\ast_2}_v$ which together satisfy the stronger property $x^{\varepsilon^\ast_2}_v.s^\ast ~<~ f.x^{\varepsilon^\ast_2}_v.s^\ast$.%

Suppose for a contradiction that, for all $\varepsilon'_2 \leq \varepsilon_2$, every $x^{\varepsilon'_2}_v$ we could choose satisfying Properties (\ref{e1225aa}, \ref{e1225bb}) for $f^\ast_v$, that is $\Lfp.f^\ast_v - \Const{\varepsilon'_2} \leq x^{\varepsilon'_2}_v \leq \Lfp.f^\ast_v$ and $x^{\varepsilon'_2}_v \Lle f^\ast_v.x^{\varepsilon'_2}_v$, satisfied the inequality
\begin{equation}\label{e1125}
 f.x^{\varepsilon'_2}_v.s^\ast \Wide{\leq} x^{\varepsilon'_2}_v.s^\ast
\end{equation}
as well --- thus failing to have the property \Eqn{e1225bb} for $f$ at $s^\ast$ that we needed to complete our proof of \Lem{l1234}. As $\varepsilon'_2$ approaches zero%
\footnote{\label{n1123}This is where we use the strengthening of the inductive assumption, the upper bound on $x$ in (\ref{e1040a}a): recall Footnote \ref{n1122}.}
we would then have from above that we can choose a sequence of $x^{\varepsilon'_2}_v$'s approaching $\Lfp.f^\ast_v$ --- and so from the continuity%
\footnote{\label{f1107}This is where analytical --- rather than $\Max$ --- continuity of $f$ is used, since the sequence of $x^{\varepsilon'_2}_v$'s is not necessarily increasing; recall Footnote \ref{f1106}.}
of $f$ we could take limits on both sides of \Eqn{e1125}, giving
$f.(\Lfp.f^\ast_v).s^\ast \leq \Lfp.f^\ast_v.s^\ast$, and so we would have
\begin{equation}\label{e1126}
 F.v \Wide{=} f.(\Lfp.f^\ast_v).s^\ast \Wide{\leq} \Lfp.f^\ast_v.s^\ast \Wide{=} v ~,
\end{equation}
where on the left we are defining a function $F$ of $v$ for use below --- our contradiction will be achieved by considering a further property of $F$, beyond the $F.v \leq v$ that we have at \Eqn{e1126} already.

\bigskip
That property is $v^\ast \leq \Lfp.F$.
To see that we argue by cases that
\[
 f.(\Lfp.f^\ast_{\Lfp.F}) 
 \Wide{=}
 \Lfp.f^\ast_{\Lfp.F}~,
\]
which by the least-fixed-point property gives us $\Lfp.f \leq \Lfp.f^\ast_{\Lfp.F}$.
Then, applying that inequality at $s^\ast$ itself, we have $\Lfp.f.s^\ast \leq \Lfp.f^\ast_{\Lfp.F}.s^\ast$, whence
\begin{Reason}
\Step{}{
 v^\ast
}
\StepR{$=$}{definition $v^\ast$}{
 \Lfp.f.s^\ast
}
\StepR{$\leq$}{immediately above}{
 \Lfp.f^\ast_{\Lfp.F}.s^\ast
}
\StepR{$=$}{fixed point}{
 f^\ast_{\Lfp.F}.(\Lfp.f^\ast_{\Lfp.F}).s^\ast
}
\StepR{$=$}{definition: $f^\ast_x.y.s^\ast \Defs x$, for all $x,y$}{
 \Lfp.F ~,
}
\end{Reason}
as required. The two cases are
\begin{description}
\item[Case $s \neq s^\ast$:] We have
\begin{Reason}
\Step{}{
 f.(\Lfp.f^\ast_{\Lfp.F}).s
}
\StepR{$=$}{
 definition: $f^\ast_x.y.s \Defs f.y.s$, for all $x,y$ and $s \neq s^\ast$
}{
 f^\ast_{\Lfp.F}.(\Lfp.f^\ast_{\Lfp.F}).s
}
\StepR{$=$}{fixed point}{
 \Lfp.f^\ast_{\Lfp.F}.s ~.
}
\end{Reason}
\item[Case $s =s^\ast$:] We have
\begin{Reason}
\Step{}{
 f.(\Lfp.f^\ast_{\Lfp.F}).s^\ast
}
\StepR{$=$}{definition $F$}{
 F.(\Lfp.F)
}
\StepR{$=$}{fixed point}{
 \Lfp.F
}
\StepR{$=$}{definition $f^\ast_\cdot$}{
 f^\ast_{\Lfp.F}.(\Lfp.f^\ast_{\Lfp.F}).s^\ast
}
\StepR{$=$}{fixed point}{
 \Lfp.f^\ast_{\Lfp.F}.s^\ast ~.
}
\end{Reason}
\end{description}
That establishes the equality we used above, and completes the demonstration that $v^\ast \leq \Lfp.F$.

\bigskip
Now from that and $v < v^\ast$ we have immediately%
\footnote{\label{n1018}This is where the strictness is used: recall Footnote \ref{n1004}.}
that $v < \Lfp.F$ also and, since $F$ is monotonic (it is constructed from monotonic pieces), by the least-fixed-point property we have $F.v \not\leq v$ --- which contradicts \Eqn{e1126}. Therefore our assumption  must fail: there must be some $0 < \varepsilon^\ast_2 \leq \varepsilon_2$ for which not all choices of $x^{\varepsilon^\ast_2}_v$ satisfying \Eqn{e1225aa} satisfy \Eqn{e1125} as well --- that is, at least one will satisfy \Eqn{e1225bb} at $s^\ast$. That is the value we take.
\end{proof}
\end{lemma}

\section{Proof of \Lem{l1228} from \App{s1650}}\label{a1250}

We prove in several stages that if $f$ is almost-linear then $f$ is $\OK$ in each argument
separately, beginning with some preliminary lemmas.

\begin{lemma}\label{l1236} {\rm Almost-feasibility} --- \quad
Let transformer $f$ be almost-linear, and suppose for some state $s$ that $f.0.y.\cdots.z.s = 0$, where \emph{wlog} we concentrate on the first argument of $f$. Then for any expectation $x$ we have $f.x.y.\cdots.z.s \leq \CompNY{\Max}{s\In S}{x.s}$.
\begin{proof}
This is clear from the explicit form (refer \Def{d1156}) that almost-linear transformers
take: if $f.0.y.\cdots.z.s$ is zero, then all non-$x$ terms $w.s, h.y.s, \cdots, i.z.s$ in $f$ must be
zero --- that is, for those values $y,\cdots, z, s$ we have $f.x.y.\cdots.z.s = g.x.s$ for some one-bounded linear transformer $g$, from which property of $g$ we have $g.x.s \leq \CompNY{\Max}{s\In S}{x.s}$.
\end{proof}
\end{lemma}

From now on we will fix the non-$x$ arguments of $f$, and omit them for brevity.

\begin{lemma}\label{l1244} {\rm Stationary zeroes} --- \quad
Let transformer $f$ be almost-linear, and define
its kernel $K$ to be those states on which its
fixed-point is zero: that is,
$K \Defs \SetCompYN{s\In S}{\mu.f.s = 0}$.
		
Then in effect the probabilistic game $f$ cannot escape from $K$: for any
state $k$ in $K$ and expectation $x$, we have
\[		
 f.x.k   \Wide{=}   f.(x\Chop K).k ~,
\]			
where $(x\Chop K).s \Defs \Cond{x.s}{s {\in} K}{0}$ .
		
\begin{proof}
Because $\mu.f$ is non-zero everywhere outside $K$, and our state space $S$ is finite,
there is an $\varepsilon > 0$ so that $\varepsilon \Times  x\Chop(S-K) ~\leq~   \mu.f$ whence, for $k$ in $K$, we have
\[
 f.(\varepsilon \Times x\Chop(S-K)).k
 \Wide{\leq}
 f.(\mu.f).k
 \Wide{=}
 \mu.f.k
 \Wide{=}0 ~.
\]	
From that, the almost-linearity of $f$ and that $\varepsilon > 0$ we have
$f.(x\Chop (S-K)).k = 0$.
Again using almost-linearity, we continue
\begin{Reason}
\Step{}{
 f.x.k
}
\Step{$=$}{
 f.(x\Chop K + x\Chop(S-K)).k
}
\StepR{$\leq$}{almost-linearity}{
 f.(x\Chop K).k \Wide{+} f.(x\Chop (S-K)).k
}
\StepR{$=$}{$f.(x\Chop (S-K)).k = 0$, shown above}{
 f.(x\Chop K).k ~.
}
\end{Reason}
		
The opposite inequality is immediate from monotonicity.
\end{proof}
\end{lemma}

\begin{lemma}\label{l1304} {\rm Almost-linear is almost \OK} --- \quad 
Let transformer $f$ be almost-linear, 
and suppose for some expectation $x$ that $x \Lle f.x$~. Then for all $k$ in the
kernel $K$ of $f$ we have $x.k = 0$.

\begin{proof}
If $x$ is not zero on $K$ then $K$ must be non-empty and there must be a state $k^\ast$ in $K$ at which $x$ attains a non-zero maximum $\CompNY{\Max}{k\In K}{x.k}$. Then because $x \Lle f.x$ we have
$x.k^\ast < f.x.k^\ast = f.(x\Chop K).k^\ast$ from \Lem{l1244}.

Now $f.0.k^\ast \leq f.(\mu.f).k^\ast = \mu.f.k^\ast = 0$, since $k^\ast \in K$,so that from \Lem{l1236} we have as well that
$f.(x\Chop K).k^\ast \leq \CompNY{\Max}{k\In K}{x.k}$~.
Taken together with the above, that gives $x.k^\ast < \CompNY{\Max}{k\In K}{x.k}$, contradicting the
choice of $k^\ast$.
\end{proof}
\end{lemma}

\bigskip
We can now proceed with the proof of \Lem{l1228}; we assume $x \Lle f.x$.

First choose a real scalar $\varepsilon > 0$ so that
\[
 \frac{\mu.f.s}{x.s} \Wide{\geq} \frac{\varepsilon}{\varepsilon+1}
 \quad \textrm{for all $s$ with $x.s \neq 0$,}
\]
which is possible because $S$ is finite, and note that $(\varepsilon +1)(\mu.f)
- \varepsilon x \geq 0$ then holds for all states --- since when $\mu.f.s = 0$ we have from \Lem{l1304} that  $x.s = 0$ as well. In fact we can decrease $\varepsilon$ still further, if necessary, to achieve
\[
 0 \Wide{\leq}
 (\varepsilon +1)(\mu.f) - \varepsilon x
 \Wide{\leq}
 1 ~,
\]
sufficient to use the expression as a one-bounded expectation in the argument below.
	
Now because $f$ is almost-linear, as a function of $x$ it is of the form $w + g.x$ for fixed expectation $w$ and linear $g$.%
\footnote{We absorb the fixed contributions $h.y,\cdots,i.z$ of other arguments into $w$.}
Applying $f$ to our expression above, and using linearity of $g$, we have
\begin{Reason}
\Step{}{
 f.(~(\varepsilon+1)\mu.f ~-~  \varepsilon x~)
}
\Step{$=$}{
 w
 \Wide{+}
 g.(~(\varepsilon+1)\mu.f ~-~  \varepsilon x~)
}
\StepR{$=$}{$g$ linear}{
 w
 \Wide{+}
 g.(~(\varepsilon+1)\mu.f~)
 \Wide{-}
 g.(\varepsilon x)
}
\Step{$=$}{
 w
 \Wide{+}
 (\varepsilon+1)(g.(\mu.f))
 \Wide{-}
 \varepsilon (g.x)
}
\Step{$=$}{
 (\varepsilon+1)w
 ~+~
 (\varepsilon+1)(g.(\mu.f))
 \Wide{-}
 \varepsilon w
 ~-~
 \varepsilon (g.x)
}
\Step{$=$}{
 (\varepsilon+1)(f.(\mu.f))
 \Wide{-}
 \varepsilon (f.x)
}
\StepR{$\leq$}{$x \leq f.x$, because $x \Lle f.x$}{
 (\varepsilon+1)(\mu.f)
 \Wide{-}
 \varepsilon x ~,
}
\end{Reason}
so showing that 
$(\varepsilon+1)(\mu.f) -  \varepsilon x$
is a pre-least-fixed-point of $f$.

Thus by the least-fixed-point property we have
\[
 \mu.f
 \Wide{\leq}
 (\varepsilon+1)(\mu.f) ~-~  \varepsilon x~,
\]	
whence by arithmetic (rearranging, dividing by $\varepsilon > 0$) we have
$x \leq \mu.f$~.

\bigskip			
Thus we have proved that all almost-linear transformers are \OK.

\end{document}